\documentclass{elsarticle}
\usepackage{natbib}
\usepackage{geometry}
\usepackage{hyperref}
\usepackage{graphicx} 
\usepackage{amsfonts}
\usepackage{amssymb}
\usepackage{mathtools}
\usepackage{mathrsfs}
\usepackage{txfonts}
\usepackage{dsfont}
\usepackage{pifont}
\usepackage{subcaption}
\usepackage{booktabs}
\usepackage{xcolor}
\usepackage{tabularx}
\usepackage{multirow}
\usepackage{multicol}
\usepackage{cleveref}
\usepackage{float}
\usepackage{arydshln}
\usepackage{tikz,pgfplots}
\pgfplotsset{compat=1.18}
\usepackage{bm}   
\usepackage{bbm}

\title{A multi-dimension and high-granularity equity measurement for transportation services through accessibility and reliability}
\author[1]{Mengke Ma\fnref{contrib}}
\ead{mm11392@nyu.edu}
\author[1]{Zilin Bian\fnref{contrib}\corref{poc}}
\ead{zb536@nyu.edu}
\author[1]{Jingqin Gao}
\ead{jingqin.gao@nyu.edu}
\author[1]{Hai Yang}
\ead{hy1236@nyu.edu}
\author[1]{Joseph Chow}
\ead{jc7373@nyu.edu}
\author[1]{Kaan Ozbay}
\ead{kaan.ozbay@nyu.edu}

\fntext[contrib]{Authors contributed equally}
\cortext[poc]{Corresponding author: Zilin Bian, email: zb536@nyu.edu}

\address[1]{Department of Civil and Urban Engineering, New York University}

\begin{document}
\begin{abstract}
Transportation equity research has traditionally emphasized service accessibility and destination reachability while often overlooking the critical aspects of service quality, such as infrequent schedules or overcrowded vehicles. This oversight can lead to a skewed understanding of equity, as high accessibility does not guarantee high-quality service. Addressing this gap, we propose a transportation equity index called the multi-dimensional, high-granularity (MDHG) index. Such an index considers service accessibility and quality alongside population demographics. This approach ensures that areas with high accessibility but low service quality are recognized as inequitable. The MDHG Index addresses service performance by incorporating performance data with temporal variations based on actual trip data, thus offering a more nuanced view of transportation equity that reflects the real-world experiences of service users. Furthermore, to effectively identify and address the needs at the user level, we need to use a highly granular population dataset. Due to the low granularity of census and other open-source datasets, we opted to use a highly granular synthetic dataset. To test out the MDHG Index, we coupled a highly granular synthetic population dataset with data from New York City’s Citi Bike expansion to use as a case study to assess changes in accessibility and service quality before and after the expansion. The MDHG approach effectively identified areas that improved post-expansion and highlighted those requiring further enhancement, thus showing the effectiveness of the index in targeted improvements for transportation equity. 
\end{abstract}



\maketitle
\section{Introduction}
Agencies of various levels of jurisdictions have been incorporating equity aspects in the transportation planning processes. While the number of available resources and tools to decision-makers has steadily risen in recent years, many focus on screening areas suitable for potential equity-focused transportation projects instead of evaluating a proposed project or the impact of a completed project \cite{Gao}. As a result, there needs to be more readily available frameworks for project evaluation. From an academic perspective, transportation equity research has traditionally focused on service accessibility and destination reachability. Levinson \cite{Levinson} found connections between shorter commute times with residences in job-rich areas and job opportunities in housing-rich areas. Dong et al. \cite{Dong} discussed the necessity to consider the accessibility to various activity-based destinations to reflect the change in accessibility according to different activities. Martens et al. \cite{2022} recently proposed a framework that shifts the focus from accessibility disparity to insufficiency. However, existing frameworks often overlook the critical aspect of service quality and performance. This oversight can lead to a skewed understanding of equity, as good accessibility does not guarantee high-quality service. For instance, services and destinations might be within reach, but if the service quality is poor—characterized by infrequent schedules, delays, or overcrowded vehicles during certain periods —the time-dependent level of service can be low, undermining the equity perspective. A prime example of this observation is the recent meltdowns in the New Jersey Transit (NJ Transit) system, where rush hour train services came to a complete halt several times due to downed power cables \cite{McGeehan}. The level of service is dynamic \cite{Jonsson} and needs to be incorporated into service project evaluations – an important observation, especially for decision-makers, to consider when addressing transportation equity issues. In short, there is a need to expand the dimensionality of equity indices from purely spatial to spatial-temporal. 

Furthermore, existing frameworks often rely on traditional census data aggregated at geographical unit levels, such as census tracts (e.g., \cite{CensusBureau}). This approach often leads to the creation of target populations with limited attribute combinations due to the reliance on pre-aggregated, publicly available demographic data. However, to better target specific populations with greater flexibility and diversity, it is crucial to incorporate higher-resolution socio-demographic data. Such detailed data would enable policymakers to identify and address the needs of specific populations more accurately and effectively. Therefore, there exists an imminent need for demographic data of higher granularity to improve the targeting of specific populations. 

\begin{figure}[H]
    \centering
    \includegraphics[width=0.9\linewidth]{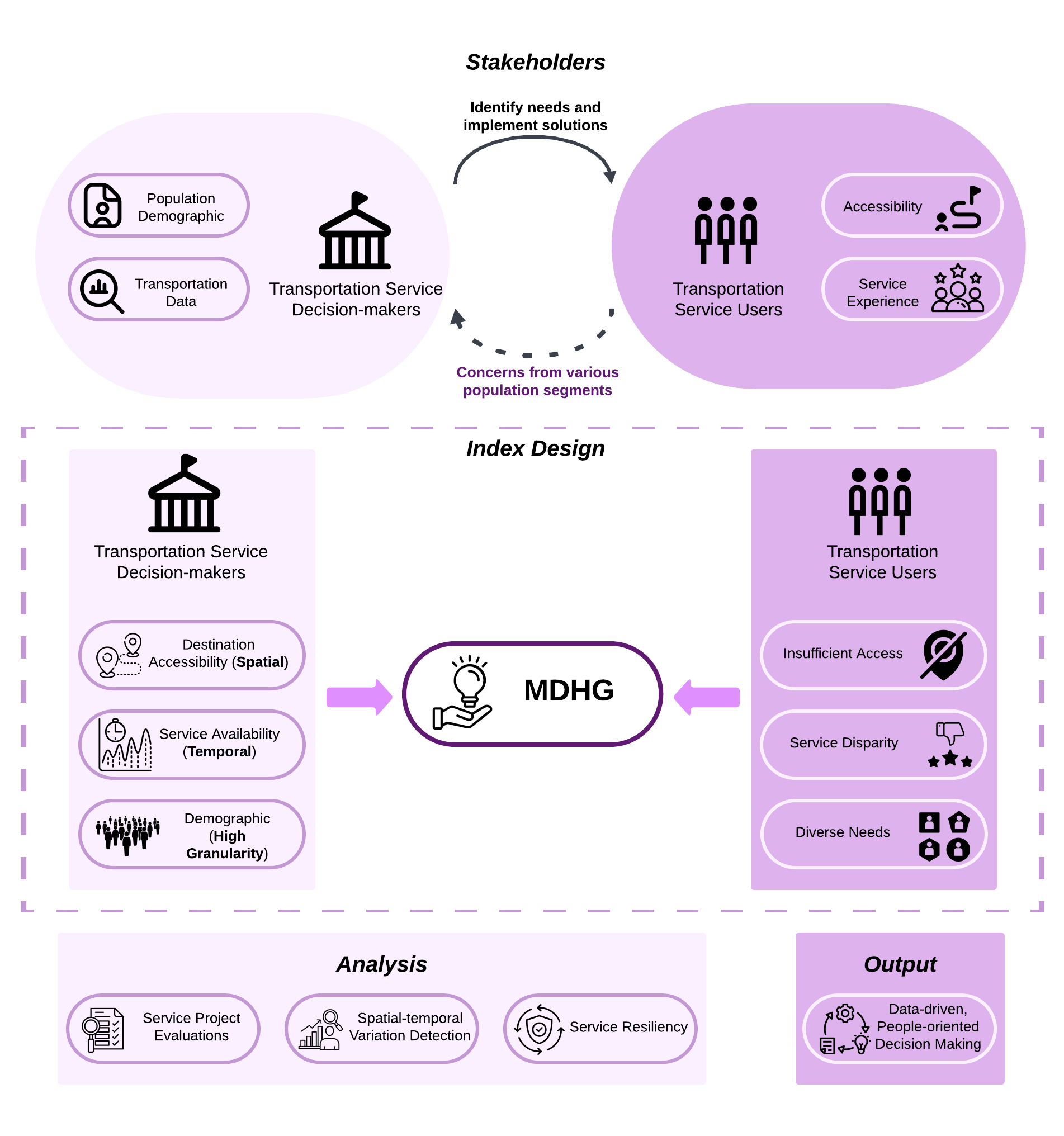}
    \caption{Overview of MDHG equity index}
    \label{fig:concept-MDHG}
\end{figure}

To address these needs and gaps, we propose a people-oriented, multi-dimensional, high-granularity (MDHG) framework (\Cref{fig:concept-MDHG}) to generate an equity index for service-related transportation projects that combines a modified accessibility insufficiency approach (\(I_{access}\)) with another metric, service performance $z$, which varies hourly based on the recorded trips throughout the day for each region in the study area. We then apply demographic overlays from a synthetic population dataset \cite{He} to show the number of targeted subgroups, such as income and ethnic-disadvantaged communities, to highlight socio-economically burdened areas that experience inadequate access or unsatisfactory service quality. Compared to the publicly available person-level Census microdata datasets (e.g., PUMS), the generated synthetic dataset fills in the gaps of missing samples. The person-level synthetic population data allows decision-makers to select different target population groups when evaluating a project, therefore enabling them to address the needs based on specific populations. 

To illustrate the proposed methodology, we conducted a case study on Citi Bike shared micro-mobility in New York City (NYC). This study analyzes the impact of Citi Bike's expansion by comparing data from two distinct periods: August 2018, representing the pre-expansion phase following the completion of Phase 2 with record trip numbers, and April 2024, which reflects the post-expansion period after the late 2023 expansion. We aim to address three main research questions using this case study: 1) how the expansion changes the destination accessibility for each neighborhood, 2) how service quality impacts this destination accessibility, and 3) how service quality and destination accessibility affect different population segments. 

In summary, the proposed framework brings the following three main contributions to the field. 

\begin{itemize}
    \item We introduce an innovative MDHG framework and a performance-incorporated accessibility index for project evaluation. Our proposed approach addresses a critical research gap by including a dynamic service availability metric in the static spatial aspect of destination accessibility.
    \item The proposed framework integrates high-granularity population segment data into the developed equity index, enabling more fine-grained analysis to evaluate transportation project outcomes based on target populations. This practical application of the research makes it highly relevant to transportation planners, decision-makers, and researchers involved in transportation equity and urban planning.
    \item We apply this new framework of a performance-incorporated accessibility index to Citi Bike shared micro-mobility in NYC. We evaluate this project by comparing the equity metrics before and after the expansion. Our findings generate several policy implications that can enhance the effectiveness and reach of shared micro-mobility services. For example, the performance metric highlights the temporal variations of service availability, and, therefore, should be included by decision-makers when evaluating service projects. 
\end{itemize}

\section{Literature Review}
We first searched for existing transportation equity tools used by decision-makers. We reviewed 12 sources in the United States, summarized in \Cref{tab:literature}, including MPOs/ local agencies, the Federal government, and third-party entities. The MPO/local agency-level resources and tools show that most of them were built for project screening purposes, and not all screening tools have an evaluation counterpart (e.g., \cite{Source4} \cite{Source7}). In addition, some screening tools for transportation equity do not use transportation-related metrics (\cite{Source4} \cite{Source8}). Although some screening tools use similar metrics (e.g., the percentage of low-income populations \cite{Source2} \cite{Source3} \cite{Source6}), the flagging thresholds may vary to reflect the local living costs or demographics. MPO/local agency-level evaluation tools (e.g., \cite{Source1} \cite{Source5}) also have different purposes, mainly covering equity needs assessment and identifying potential impacts of proposed projects, but generally lack metrics and guidelines for post-implementation progress evaluation. While federal-level screening tools (\cite{Source9} \cite{Source10}) are useful, the nature of their wide range of indicators and selection methods indicates that their area selection may not be the most appropriate for the specific objectives of a transportation project in a local area. For example, these tools may help easily target a low-income census tract with high housing costs. Still, they may not necessarily show transportation disadvantages, such as a lack of public transit access. Although some tools compare the study area to the state and the rest of the nation, they may overlook some individual indicators. For example, a study area may not be categorized as experiencing transportation disadvantages due to the composite score of this topic not meeting the threshold in the state and nationwide comparison. Still, the frequencies of transit services may be lower than the selected area’s average and need to be addressed. In addition, equity performance evaluation measures identified for the national level may need to be further developed and adapted to be applicable at a more local level. The two third-party tools (\cite{Source11} \cite{Source12}) provided insights on screening areas and needs but require more details for post-project evaluations. To summarize, the reviewed existing resources and tools in the United States indicate that while screening/prioritization tools are more mature, more evaluation tools, such as an equity index, and guidelines need to be developed for post-implementation to evaluate the progress made by equity-centered improvement goals of the projects. 

Considering the observation above, we looked for journal articles on developing a transportation equity index. Beiler and Mohammed \cite{2016} proposed a transportation justice framework called the Transportation Justice Threshold Index Framework (TJTIF) that calculates a composite score based on demographics, socio-economics, and transportation/land use factors to evaluate whether an area needs transportation equity-related projects. This work aimed to provide a streamlined screening method applicable across transportation agencies since many currently have different criteria for selecting metrics to flag an equity-focus area. While this work provides a comprehensive screening method, it does not include a post-project evaluation component. More recently, Collins et al. \cite{2023} developed a different framework titled the Transit Equity Index (TEI), comprised of a transit service index, socially disadvantaged index, and race index. While this work provides a novel method of calculating an equity index for transit services and may be used for post-project evaluation when using data from multiple years, the methodology does not include metrics relevant to destination accessibility, and may need to be generalized to be applied to transportation projects other than transit. 

Shifting the gear to look for quantitative literature that includes destination reachability, studies that have sought to quantify impacts of operations on accessibility have resorted to metrics like Gini coefficient (e.g. \cite{Delbosc}) or similar metrics that measure disparities (e.g. \cite{Ferguson}, \cite{Dai}) as opposed to ensuring justice for underserved groups. In the case of bikeshare, prior research on measuring equity in accessibility has not considered time dimension in service operations (\cite{Desjardins}, \cite{Jin}). 

 The insufficient access work by Martens et al. \cite{2022} provides a distinct perspective that shifts from destination accessibility disparity to insufficiency, in which the same accessibility threshold applies across the entire population to identify areas below the threshold, thus creating a new benchmark for equity priority areas focused more on ensuring justice. \cite{Ren} derived a mode choice-based measure of equity from this insufficiency metric and applied it to Replica data across New York State. While the work by Martens et al. addresses destination accessibility insufficiency, the framework does not include any metric on the quality of service, which may result in a skewed understanding of equity in scenarios where destination accessibility is excellent but service is unsatisfactory. 

\begin{table}[H]
\centering
\begin{tabular}{ccccc}
\toprule
\multirow{2}{*} {Resource/}& {Jurisdiction} & Screening/ Prioritization  & Includes Transportation- & Applicable to Post-\\
       Tools& &With Quantitative Metrics & Related Metrics&Project Evaluation \\
\midrule
\cite{Source4}& MPO & Yes  & No & No \\
\midrule
\cite{Source7}& City & Yes  & No & No \\
\midrule
\cite{Source8}& State & Yes  & No & No \\
\midrule
\cite{Source2}& City & Yes  & Yes & No  \\
\midrule
\multirow{2}{*} {\cite{Source3}}& City & Yes  & Yes & Yes, due to \\
&&&&pre-defined goals\\
\midrule
\cite{Source6}& MPO & Yes  & Yes & No \\
\midrule
\cite{Source1}& City & Yes  & Yes & No  \\
\midrule
\cite{Source5}& MPO & Yes  & Yes & No \\
\midrule
\multirow{3}{*} {\cite{Source9}}& Federal & Yes  & No & No, but Justice40/ \\
&&&&USDOT have pre-\\
&&&&defined goals\\
\midrule
\multirow{3}{*} {\cite{Source10}}& Federal & Yes  & Yes & No, but Justice40/ \\
&&&&USDOT have pre-\\
&&&&defined goals\\
\midrule
\cite{Source11}& Third-Party & Yes  & Yes & No \\
\midrule
\multirow{3}{*} {\cite{Source12}}& Third-Party & Yes  & Yes & Has potential  \\
&&&&due to frequent\\
&&&&data updates\\
\bottomrule
\end{tabular}
\caption{\small Key takeaways from reviewing 12 existing equity resources/ tools}
\label{tab:literature}
\vspace{-0.5cm}
\end{table}

In summary, the review of existing agency tools and academic literature reveals a lack of quantitative post-project evaluation methods and the integration of destination accessibility with service quality in transportation equity analysis. To address these gaps, we develop a multidimensional, high-granularity framework to quantify equity for service-related transportation projects. 

\section{Methodology}
This section outlines the design methodology for the proposed equity index and the data preparation process for the case study. The literature review revealed the necessity of a quantitative metric for service performance ($z$). Consequently, for index development, we included a component score representing service performance disparity ($D_{sp}$), in addition to the score for destination accessibility insufficiency ($I_{access}$). These components were then combined to create a composite score ($ZI_{access}$) for ease of application by agencies. For the case study application, we calculated each score component for each census tract before and after the Citi Bike expansion. To ensure comparability, we restricted our analysis to the 1117 tracts with a bike rack in either period. All notations involved are presented in \Cref{tab:notations}. 

\begin{table}[H]
    \centering
    \begin{tabular}{cc}
    \toprule
        $i$ & Each study unit \\
        $h$ &  Each temporal timestamp\\
        $A$ & Available service (numeric)\\
        $C$ & Total service capacity (numeric)\\
        $\zeta$ & Service availability (ratio)\\
        $H$ & Reversed percentile rank \\
        $z$ & Performance metric\\
        $\eta$ & Percentile rank\\
        $n$ & Population count\\
        $pm$ & Desired performance metric(s) to include\\
        $D_{sp}$ & Service performance disparity metric \\
        $k$ &  Neighborhood mean of area of destination reachability\\
        $x$ & Area of destination reachability\\
        $I_{access}$ & Destination access insufficiency metric\\
        $ZI_{access}$ & Performance-integrated destination access insufficiency metric \\
        $R$ & Recovery time to a service availability threshold\\
        $l$ & Level of service availability threshold\\
        $SR$ & Service recovery rate metric\\
    \bottomrule
    \end{tabular}
    \caption{Notation of variables, in sequential order}
    \label{tab:notations}
\end{table}

\subsection{Equity Index}
\begin{figure}[H]
    \centering
    \includegraphics[width=0.88\linewidth]{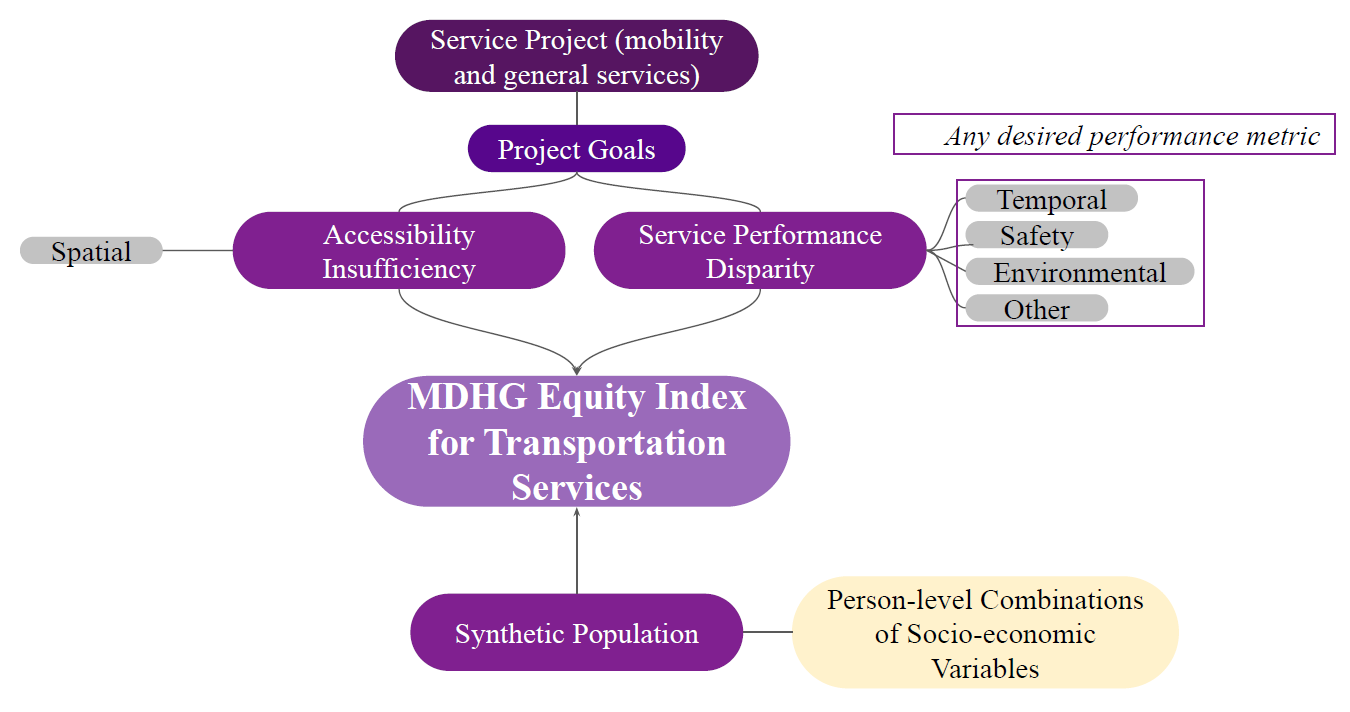}
    \caption{Overview of the equity index development}
    \label{fig:equity-develop}
\end{figure}

We present the workflow overview of the equity index development in \Cref{fig:equity-develop}. 

\subsubsection{\textit{Performance Metric, $z$}}
We began by designing the performance metric $z$ to reflect the nature of temporal and spatial variations of service availabilities. For different transportation projects, $z$ should be adaptable to reflect their specific service. For example, in a subway project, the performance metric may be a combination of average trip delays, service frequency, and crowdedness. Another example of electric vehicle charging stations may use the number of available chargers at a given time and charging speed as the performance metrics. 
\begin{equation}
\zeta_{i, h} = \frac{A_{i, h}}{C_{i, h}}
\label{eq:zraw}
\end{equation}
\begin{equation}
z_{i, h} = H(\zeta_{i, h})
\label{eq:zrank}
\end{equation}
\\
A higher $z$ value indicates a lower $\zeta$ value, which represents low service availability.

\subsubsection{\textit{Service Performance Disparity, $D_{sp}$}}
After obtaining the $z$ values, we designed a service performance disparity score, $D_{sp}$, that ranges from 0 to 1, where a value closer to 1 would indicate that more people are experiencing unsatisfactory service performance qualities.
\begin{equation}
D_{sp_{i,h}} = minmax(pm_{i,h}\cdot\eta(n_{i}))
\label{eq:dsp}
\end{equation}
\\
In short, $D_{sp}$ is based on the min-max normalized product of the ranked performance metric(s) ($pm_{i, h}$, effectively $z$ in the case study) and the rank of the population ($n$) for each study unit.

\subsubsection{\textit{Destination Access Insufficiency, $I_{access}$}}
Inspired by Martens et al. \cite{2022}, we defined our destination access insufficiency score as the following. A score closer to 1 indicates that more people are experiencing insufficient destination access. 
\begin{equation}
I_{access_{i}} = minmax(n_{i}\cdot(\frac{k_{i}-x_{i}}{k_{i}})^2)
\label{eq:iacc}
\end{equation}
\\
Where $n_{i}$ is the population in an individual study unit, $x_{i}$ is the destination accessibility in that study unit, and $k_{i}$ is the neighborhood average for accessibility. Accessibility can be defined flexibly based on the evaluation needs. For example, it can represent the number of jobs reachable or the area of activity zones reachable from each neighborhood. We separately took the log-normalized values of the $n$ and $(\frac{k-x}{k})^2$ before multiplying to ensure that both parts contribute on a similar scale.

\subsubsection{\textit{Multi-Dimensional High-Granularity Equity Index for Transportation Services, $ZI_{access}$}}
Since we identified combining service quality with accessibility insufficiency as a research gap, we combined the two components using the following equation. In other words, $ZI_{access}$ represents the proposed MDHG equity index. A score closer to 1 indicates that more people are experiencing insufficient access and unsatisfactory service quality. 
\begin{equation}
ZI_{access_{i, h}} = minmax(z_{i,h}\cdot\eta(n_{i}\cdot(\frac{k_{i}-x_{i}}{k_{i}})^2))
\label{eq:ziacc}
\end{equation}
\\
We decided to exclude the $\eta(n)$ element from $D_{sp}$ since the $I_{access}$ configuration already captures the population. In short, $ZI_{access_{i, h}}$ is a combination of modified $I_{access}$ and $z$, where the $z$ component adds a spatial-temporal dimension to the spatially variating \(I_{access}\), reflecting the accessibility's nature of temporal-dependency based on service availability. We applied the same log-normalization to $n$ and $(\frac{k-x}{k})^2$. 

\subsubsection{\textit{Service Recovery Rate ($SR$)}}
We defined a new metric, Service Recovery ($SR$), to spatially inspect the temporal variations of service availability. This rate can be used as an approximation indicating service resilience.
\begin{equation}
SR =  \int_{l_1}^{l_2} \mathbf{R_l}\,dl \
\label{eq:SR}
\end{equation}
\\
We focused on the metric $\zeta$, where given a specific threshold for $\zeta$, we calculated the temporal duration it took for the $\zeta$ values in the study unit to recover ($R$) to the threshold level ($l$). The output of $SR$ is twofold. First, we created a choropleth map showcasing each region's recovery time to each threshold level. Then, we calculated the vector integral in each study unit, as shown in \Cref{eq:SR}, for different recovery times based on various thresholds. A greater integrated value indicates an overall longer recovery time for each threshold, suggesting that service availability in that study unit is often below the selected thresholds.

\subsection{Case Study Data}
\subsubsection{\textit{Citi Bike Data}}
We chose Citi Bike phase 3 expansion as the case study scenario. We used the Citi Bike trip data from August 2018 to represent the period before the expansion and April 2024 to represent the period after. We applied the same filters to both datasets: 1) only keep trips that Citi Bike members completed, 2) the trip duration needed to be longer than 1 minute and shorter than 4 hours, 3) trip speed needed to be less than 650m/min (\textasciitilde 24mph), and 4) the trips had the same start and end date. We decided to restrict the trips to Citi Bike members only to focus more on non-tourist regular trips. After inspecting the distribution of the trip durations, we decided on the 1-minute and 4-hour cutoffs to remove the outliers. While Citi Bike trip data did not provide a trip distance column, we calculated the Euclidean distance based on the coordinates of the start and end stations. Since this is the shortest distance between the start and end points, the trip distance would likely be longer, thus leading to a faster trip speed. As a result, we used 650m/min as the upper limit for reality checks for casual bike riders. We also restricted trips to have the same start and end date to simplify the index calculations. 

\subsubsection{\textit{Geographical Boundaries}}
In addition to the Citi Bike trip data, we also needed the geographical boundaries for census tracts and Neighborhood Tabulation Area (NTA). We obtained the 2020 Census Tracts data from the Census Bureau \cite{tiger} and the NTA data from NYC Open Data \cite{nta}. It is important to note that Staten Island is not included in this case study since this borough currently lacks Citi Bike service. Spatially joining the filtered tracts with the station point layer by creating a 500m buffer around the border of each tract and finding the intersected stations within each buffer, we observed 1117 tracts that had a bike rack before and/ or after the expansion. To better compare the before and after results, we calculated the index components for only these 1117 tracts. We utilized the remaining tracts that never had Citi Bike access by plotting them on the final result maps. 

\subsubsection{\textit{Synthetic Population}}
While the Census Bureau extensively collects individual-level data, only a small subset of samples is available for public use due to privacy concerns, and the full census details are aggregated into marginal attributes. For example, less than one million entries of detailed individual attributes are available in the public use microdata samples (PUMS) from the 2021 5-year American Community Survey (ACS) published by the US Census Bureau in NYS, while the whole population is around 20 million in 2021 \cite{CensusBureau}. Therefore, proper population synthesis methods are required to upscale the limited sample to the full size. 

Recent approaches using deep learning, such as Generative Adversarial Networks (GANs), have overcome some challenges of dimensionality (\cite{synth_2}, \cite{synth_3}). However, GANs often lack control over the generation process, particularly in maintaining associations between individual household members (\cite{synth_4}). Conditional input directed acyclic tabular generative (ciDATGAN) builds upon previous work by introducing directed acyclic graphs (DAGs) and conditional inputs to enforce structure and limit overfitting (\cite{synth_5}). This way, the model provides enhanced control and interpretability in population synthesis. We use a hybrid model combining deterministic and ciDATGAN to enhance synthetic population quality while preserving attribute associations. This process consists of three primary steps. First, We compiled household and personal data from the 2021 American Community Survey (ACS) \cite{CensusBureau}. The dataset includes both household-level attributes (e.g., household size, income, housing type) and personal attributes (e.g., age, gender, employment status). Data is segmented by household size, and household-level attributes are linked to corresponding personal attributes. This segmentation ensures that household dynamics, such as age distribution among household members, are preserved.

\begin{figure}[H]
    \centering
    \includegraphics[width=0.8\linewidth]{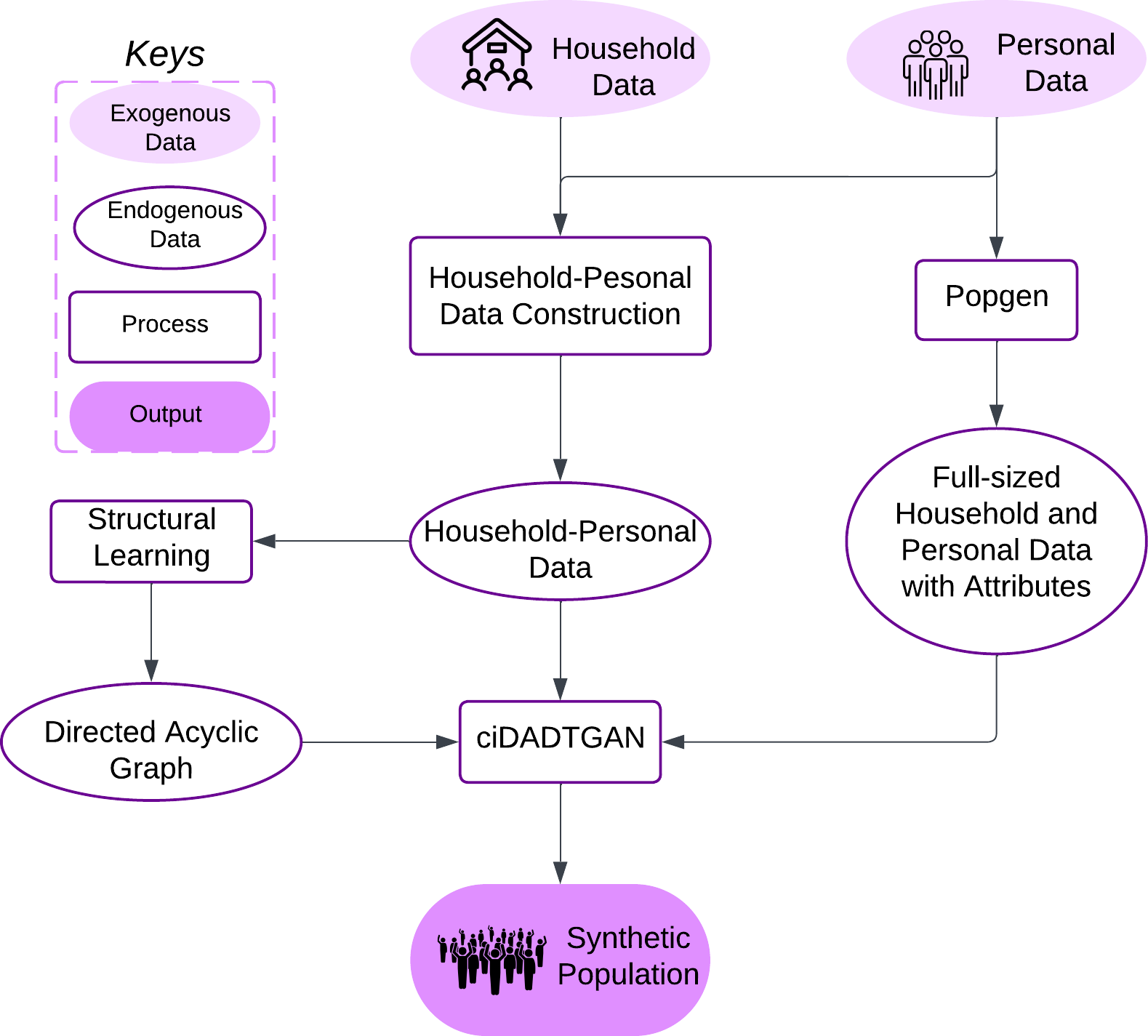}
    \caption{Population synthesis workflow}
    \label{fig:synth_pop}
\end{figure}

Our next step was to determine the most suitable DAG structure for capturing attribute dependencies. We applied six different methods based on the Bayesian network (\cite{synth_7}) to construct six choices of DAGs for selection. The involved methods include manual adjustments, self-learning, human-adjusted self-learning, and machine learning-based techniques such as Ordinary Least Squares (OLS) regression and Random Forest. Each method aimed to identify causal relationships and attribute dependencies that best reflected real-world interactions. We selected the DAG that best depicts causal relationships judged by the Akaike Information Criterion (AIC) as input for ciDATGAN, ensuring realistic associations between attributes. The last overall step was to prepare conditional inputs, often used to control the generation process and mitigate biases introduced by the training data. Conditional inputs such as geographic area, age, and gender are used to generate attributes conditioned on these foundational characteristics. Assuming the survey sample is close to the "ground truth" of the whole population for these attributes, deterministic methods are suitable for synthesizing the full-sized population with lower dimensions. We refer to the generated output as the conditional population. We began by synthesizing the conditional population across all household sizes. Then, household size-specific conditional populations can be selected for the final population synthesis. This helps ensure that the synthetic population accurately represents the diversity seen in real data, such as maintaining consistent household types across different geographic areas or ensuring that employment rates match those of specific demographic groups.
ciDATGAN is trained separately for households of different sizes, which allows for more accurate modeling of intra-household dynamics. The DAGs help in retaining attribute relationships by structuring the flow of information in the generative process, while the use of conditional inputs mitigates biases from the original data sample. This hybrid approach allowed ciDATGAN to effectively handle high-dimensional datasets while ensuring that generated outputs retained both consistency and diversity. \Cref{fig:synth_pop} illustrates the entire population synthesis process.

The framework in \Cref{fig:synth_pop} generates the full synthetic population data at the household and individual level for NYC, comprising approximately 20 million individuals and 7.5 million households. The marginals obtained from the synthetic population well match the census marginals and enable a more robust urban and transportation analysis. After only keeping person-level and household-level data from the non-Staten Island tracts, we utilized the synthetic population data for three purposes. First, we obtained the population count at the tract level $n_{i}$ to use in \Cref{eq:dsp} through \Cref{eq:ziacc}. Second, we used tract-level demographic information to further explore the spatiotemporal relation between the calculated equity index while simultaneously considering the specific demographic attributes. Third, we created customizable population overlays by mapping the household attributes to the person-level data based on the unique household number, followed by creating binary columns to indicate whether a person is disadvantaged based on values in each attribute. We consulted the guidelines in the NYC Government Poverty Measure 2020 report \cite{NYC} to indicate whether a person is under the poverty line based on family size and household income. We considered the following characteristics disadvantaged for these attributes: speaks limited English, is of non-white race, is under the poverty line, does not own a car, and is unemployed. We then calculated the tract-level population based on each binary attribute. Since the synthetic population data was at the person level, we created 31 combination binary attributes based on the individual ones. After aggregating the person-level data to the tract level, we created demographic tract overlay layers for the final result visualizations based on the 75 percentile of the population percentage for each combination. This combination approach showcased the customizability of the highly granular synthetic population data, providing valuable insights for decision-makers to address the specific needs of different population groups. We applied selected overlays to the service recovery rate (SR) result to demonstrate their usage.

While this work focuses on using the ciDATGAN-based synthetic population dataset in the MDHG framework, it is worth noting that other population datasets, such as PopGen-based data (\cite{popgen}), are also compatible with our proposed framework. In essence, depending on the desired study areas and project evaluation objectives, practitioners may choose the granularity and quality of population data based on their needs.

\subsection{Case Study Application}
\begin{figure}[H]
    \centering
    \includegraphics[width=1\linewidth]{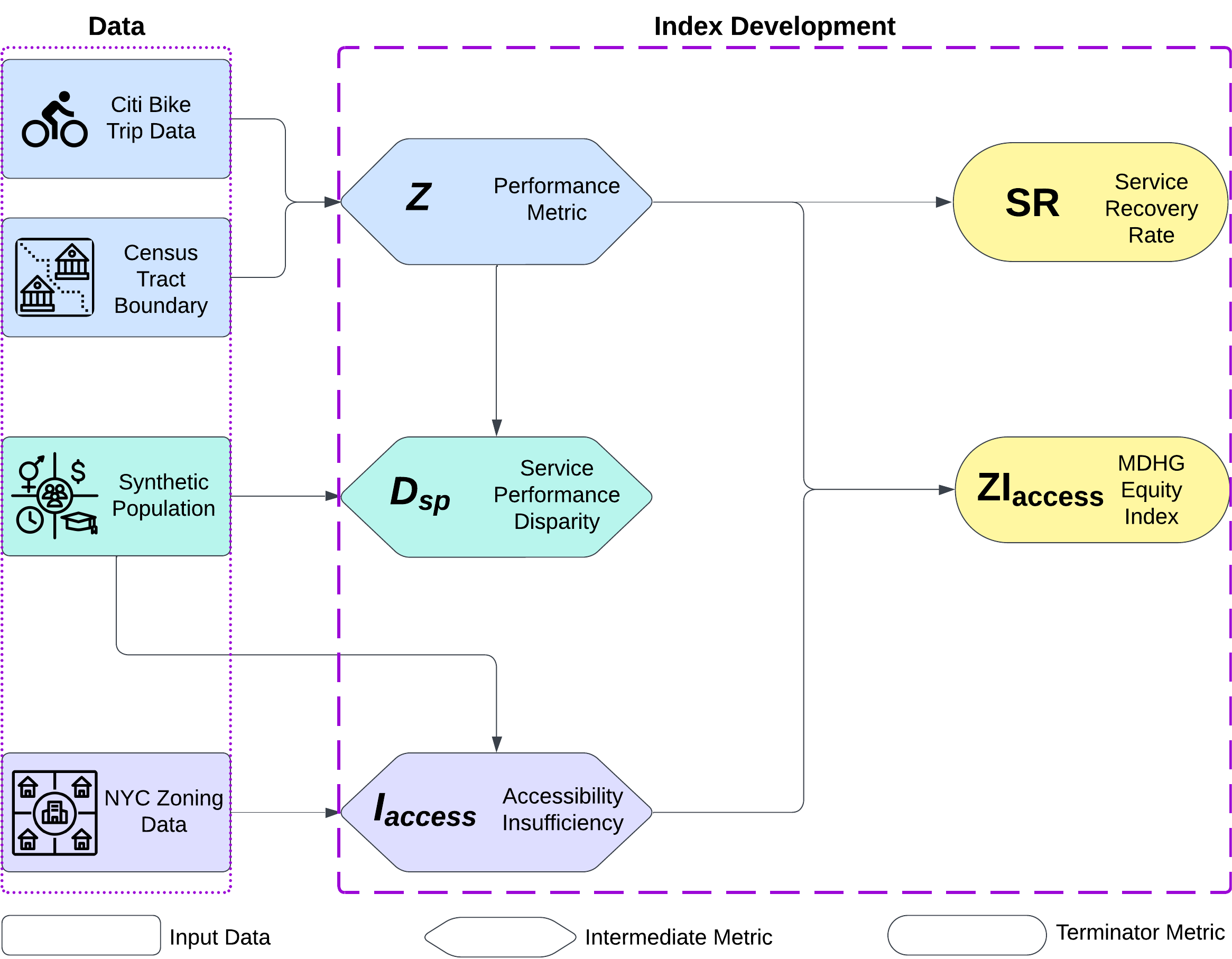}
    \caption{Overview of the case study application}
    \label{fig:case-study}
\end{figure}

\Cref{fig:case-study} presents the overview of the case study application. As discussed earlier, we aim to address three main research questions using this case study: 1) how the expansion changes the destination accessibility for each neighborhood, 2) how service quality impacts this destination accessibility, and 3) how service quality and destination accessibility affect different population segments. We answer the first question by comparing the \(I_{access}\) values before and after the Citi Bike expansion. By incorporating a temporal dimension $z$, we can observe the variations in destination accessibility based on service availability, thus answering the second question. We also separately calculate the accessibility to commercial or recreational zones based on the NYC Zoning Districts. This way, we can reflect the temporal variation to different types of destination zones. We use the synthetic population data to observe how the service availability-incorporated equity index affects population segments defined by various demographic attributes. 

\subsubsection{\textit{Performance Metric, $z$}}
\Cref{eq:zraw} and \Cref{eq:zrank} provide the calculation methods for $\zeta$ and $z$. For the Citi Bike case study presented here, we define $z$ as the reversed percentile rank of the ratio of available bikes and the total bike dock capacity for each census tract ($i$) for each hour of the day ($h$). 

After obtaining the cleaned trip data before and after the Citi Bike expansion, for each station that appeared in the dataset, we first sorted each recorded trip in an ascending order based on its timestamp. We designated an incoming trip as supply while an outbound trip as demand. A performance value representing the real-time availability by adding or subtracting 1 based on a trip’s supply or demand designation was then created. We calculated the hourly average of the performance value for each hour of the day on each day of the month and generated each station's hourly average performance value throughout the month.

As discussed earlier, the station-based trip data was spatial-joined to each census tract by creating a 500m buffer around the border of each tract and finding the intersected stations within each buffer. In other words, a bike station may fall into multiple tracts. The 500m buffer size encompasses the distance from a 5-min walk, representing the time shared-bike useres are willing to travel to access the service \cite{NACTO}. From there, we calculated the total number of available bikes at a given hour and the total bike dock capacities, and took the ratio of the two to arrive at $\zeta_{i}$ for each tract. After taking the reversed percentile rank of $\zeta_{i}$, we obtained the tract-level performance metric, $z_{i}$ before and after the Citi Bike expansion.

\subsubsection{\textit{Service Performance Disparity, $D_{sp}$}}
When applying \Cref{eq:zraw} and \Cref{eq:zrank}, we only calculated $\zeta$ and $z$ for tracts that had access to bikes due to using availability obtained from recorded trip data. As a result, we assigned a $D_{sp}$ value of 1 for tracts without service to indicate this lack. While $D_{sp}$ provides spatial-temporal observations for the service quality, we still need a component that calculates the accessibility insufficiency to represent transportation equity comprehensively.  

\subsubsection{\textit{Destination Access Insufficiency, $I_{access}$}}
Here, we use the activity area that residents from a tract can travel to using Citi Bike service as a proxy for destination accessibility (i.e., $k_i$ and $x_i$ in \Cref{eq:iacc}). We separate the zones appearing in the NYC Zoning Districts shapefile into commute (i.e., commercial and manufacturing zones) and recreational zones (commercial zones and parks). We chose these two categories to capture the most common trip purposes. We decided to include commercial zones in both categories since NYC has many mixed-used zones for office work and recreational activities, such as shopping. To obtain $x_{i}$, we first created a 2600m buffer around the border of each 1117 census tract with Citi Bike access. 2600m represents the average distance covered in shared-bike trips \cite{Li}. We then calculated the intersection area between this buffer and the two activity zones. We obtained the NTA-mean value of ${x_i}$ to obtain ${k_i}$. We applied $n\cdot(\frac{k-x}{k})^2$ only to the tracts that are below their NTA average. In summary, we calculated $I_{access}$ based on the two activity zone categories before and after the Citi Bike expansion.

\subsubsection{\textit{Performance-Integrated Destination Access Insufficiency, $ZI_{access}$}}
The $ZI_{access}$ was calculated based on \Cref{eq:ziacc}. With 24 $ZI_{access_{i,h}}$ for each activity zone category, the proposed equity index simultaneously captures spatial and temporal variations. 

\subsubsection{\textit{Service Recovery Rate ($SR$)}}
For the case study, $l$ in \Cref{eq:SR} is defined by the availability of bikes. We used 30\% to 80\% capacity as the thresholds to observe the change in recovery time for each tract. After spatially inspecting the recovery time $R$ to each threshold, we created a choropleth map of $SR$ with a few selected population overlays generated from the synthetic population data to showcase how different population segments may experience different service recovery levels. 

\section{Results and Discussion}
\subsection{\textit{Service Availability, $\zeta$}}
\begin{figure}[H]
    \centering
    \includegraphics[width=0.99\columnwidth]{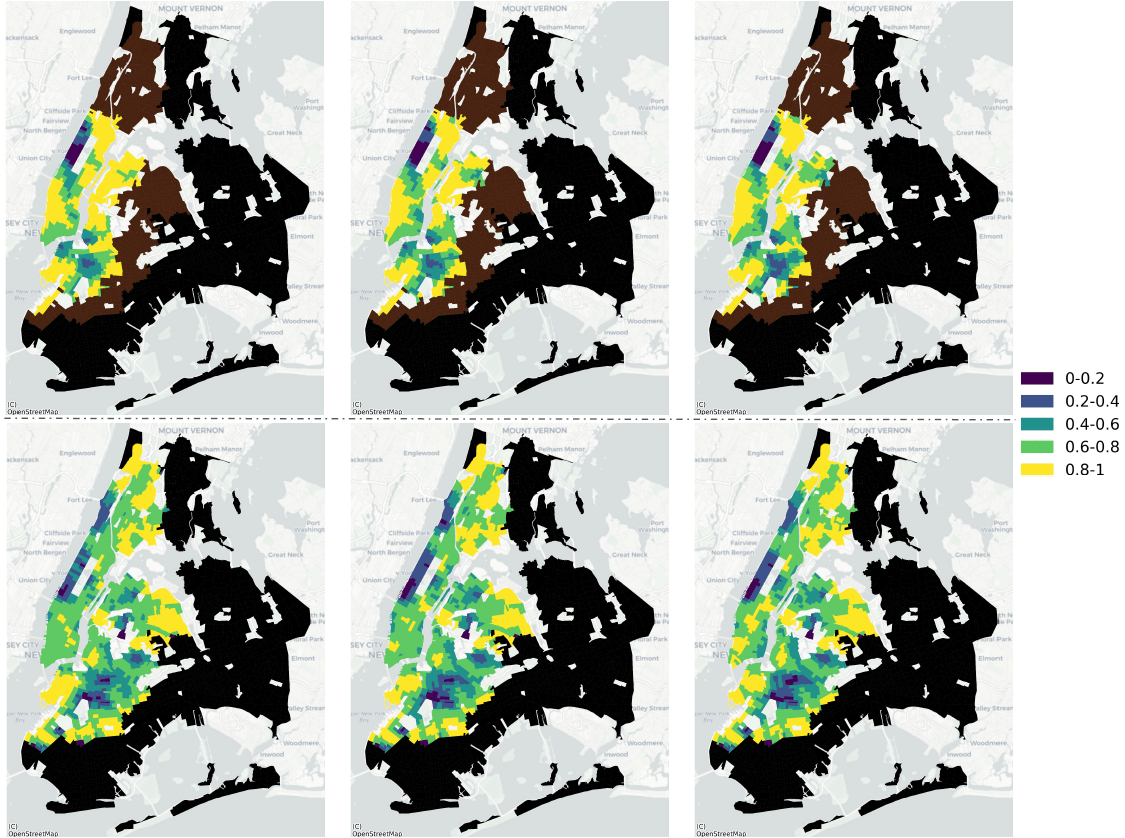}
    \caption{$\zeta$ values for hours 0, 8, 16 before (above dashed line) and after the Citi Bike expansion (below dashed line). Brown represents tracts that do not yet have access to Citi Bike services. Black represents tracts that never received Citi Bike services.}
    \label{fig:zraw_few}
\end{figure}

\begin{figure}[H]
    \centering
    \includegraphics[width=1\linewidth]{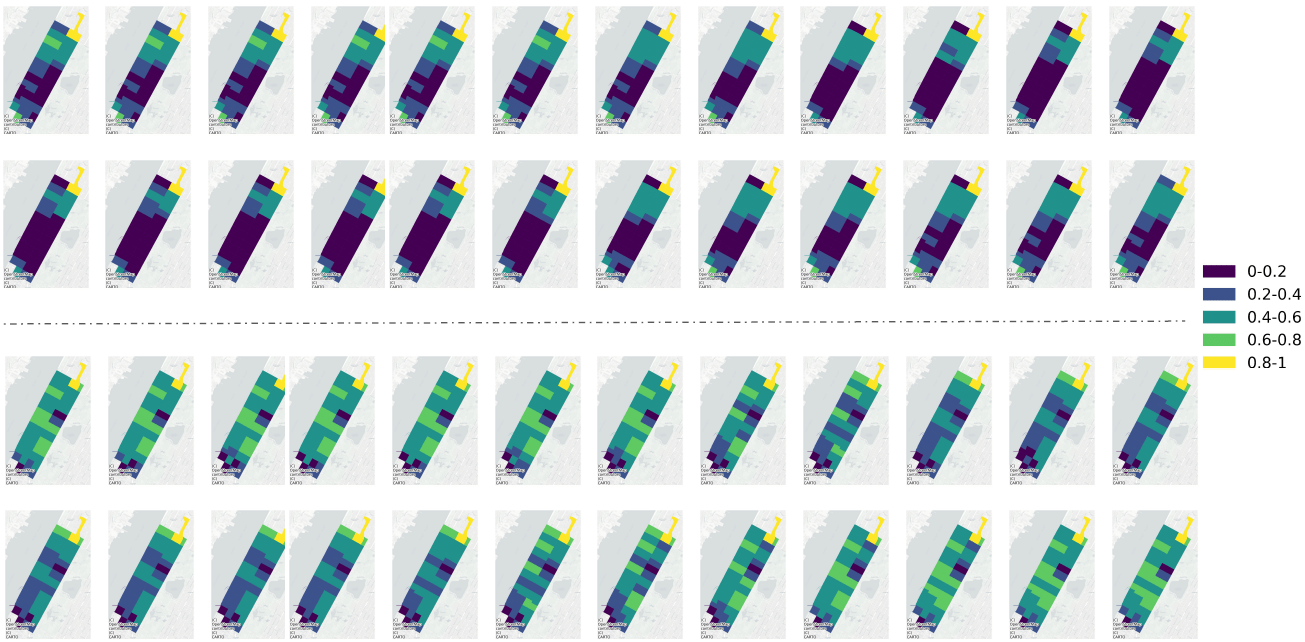}
    \caption{$\zeta$ values in UWS for all 24 hours (top row: hours 0-11; bottom row: hours 12-23), before (above dashed line) and after the Citi Bike expansion (below dashed line)}
    \label{fig:uws_all}
\end{figure}

Due to the similarities between $z$ and $\zeta$, we focus on the latter for the discussion. \Cref{fig:zraw_few} shows the $\zeta$ values for hours 0, 8, and 16. As an example, we select the Upper West Side (UWS) to display the $\zeta$ values for all hours of the day \Cref{fig:uws_all}. This shows that $\zeta$ (and also $z$) can provide a temporal dimension to the performance metric spatial distribution. 

By comparing the before and after rows in \Cref{fig:zraw_few}, we observed that most of the expansion zone tracts, especially in the Bronx, had an hourly average of $\zeta$ above 0.6, indicating that the number of bikes available to residents in these tracts consistently remained above 60\% of the total capacity. The range for $\zeta$ was between 0.107 and 1 before the expansion, while the $\zeta$ values ranged from 0.306 to 1 after the expansion. This observation suggests that the minimum service availability of bikes has improved during the expansion period. In addition, more tracts had a service availability $\zeta$ value below 0.8 after the expansion, potentially due to increased active Citi Bike membership. By looking at the 24-hour progression figures, we identified 6 to 10 AM and 5 through 8 PM as hours with high activities, likely due to commuting, by looking at the change in $\zeta$. Focusing on the UWS neighborhood (\Cref{fig:uws_all}), many tracts consistently have $\zeta$ values below 0.2, especially before the service expansion. This observation might be explained by the fact that this tract served as a popular origin that many Citi Bike users unlock bikes from this neighborhood but return them to different areas \cite{CitiBike}. This is a good example showing that while the tract has direct access to Citi Bike stations, its service availability during peak hours can be low. There was an overall increase in $\zeta$ values for this neighborhood after the expansion. One possible reason can be the expansion allowed more new origins from upper Manhattan and Bronx, and these new trips may use UWS as their destination, so the returned bikes increase the service availability in the neighborhood.

\begin{figure}[H]
    \centering
    \includegraphics[width=1\linewidth]{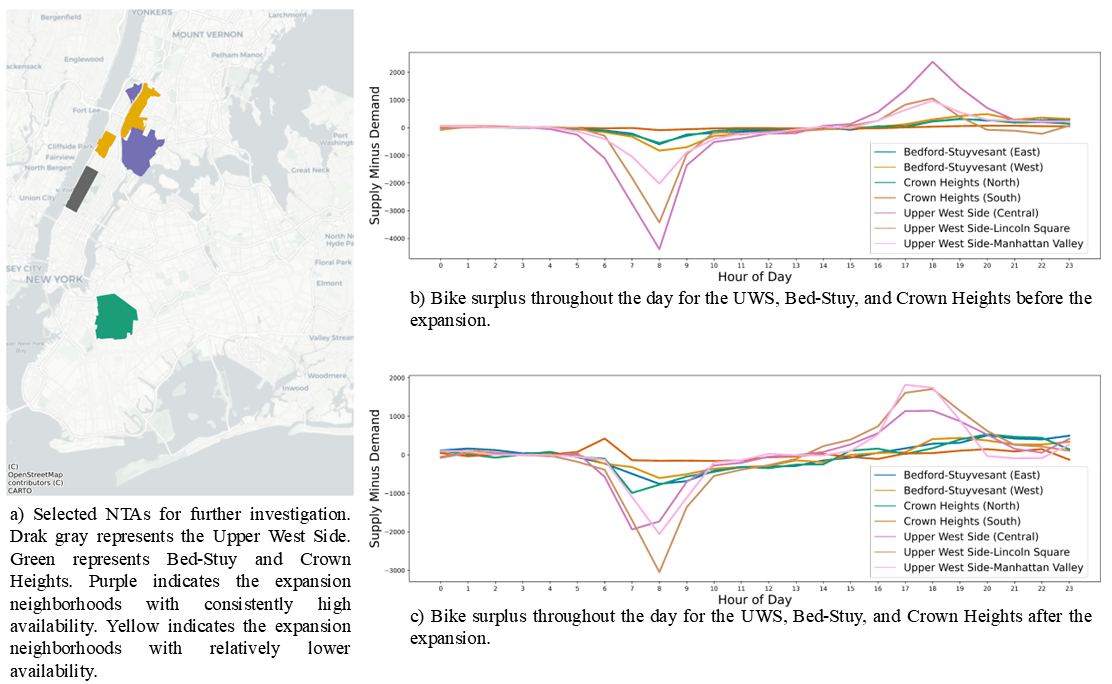}
    \caption{Selected NTAs for further analysis; bike surplus in UWS, Bed-Stuy, and Crown Heights, before and after the expansion}
    \label{fig:nta_od_bvsa}
\end{figure}

We selected 19 NTAs to further investigate the neighborhood-level (i.e., multiple tracts combined) service availability based on the supply and demand at each hour. We selected 7 neighborhoods representing the overall UWS area in Manhattan, plus Bedford-Stuyvesant (Bed-Stuy) and Crown Heights in Brooklyn, to compare the supply and demand change before and after the expansion. These three areas consistently saw relatively low service availability in both periods. The purpose of inspecting these 7 NTAs is to check whether there is a constant imbalance between supply and demand. We also selected 12 NTAs in the Bronx and upper Manhattan, places with the most disadvantaged tracts according to the population overlays, as focus areas to evaluate the supply and demand after the expansion. 6 of the expansion zone NTAs had $\zeta$ values frequently above 0.8, while the other 6 had $\zeta$ values between 0.4 and 0.6 but mostly leaning toward the latter, thus indicating their availability to be better than areas in UWS and Bed-Stuy. In addition to checking the supply and demand in these 12 NTAs, we also wanted to verify whether the bikes in areas with $\zeta$ values frequently above 0.8 are being used. \Cref{fig:nta_od_bvsa} a) shows the locations of these selected NTAs. 

We began by inspecting the surplus (i.e., the difference in supply (positive) and demand (negative)) for the 6 NTAs in the UWS, Bed-Stuy, and Crown Heights, shown in \Cref{fig:nta_od_bvsa} b) and c). \Cref{tab:od_ds} records the number of trips with their status during the 8 AM (Hour 8) and 6 PM (Hour 18) commute rushes for these 6 NTAs in the two periods.

By comparing the magnitude of surplus (vertical axis), using 8 AM as an example, in \Cref{fig:nta_od_bvsa} b) and c), we observed a prominent decrease in the negative surplus (\textasciitilde 31\%) for the UWS, while it worsened for Bed-Stuy and Crown Heights (\textasciitilde 11\% increase in the magnitude of negative surplus). \Cref{tab:od_ds} shows that bike usage increased significantly after the expansion for Bed-Stuy and Crown Heights (129\%), while the usage increase in UWS was around 26\%. Since the selected Brooklyn neighborhoods had worsened in the surplus while their ridership more than doubled, the observations suggest that these two areas should better balance supply and demand to accommodate the usage increase.

\begin{figure}[H]
    \centering
    \includegraphics[width=1\linewidth]{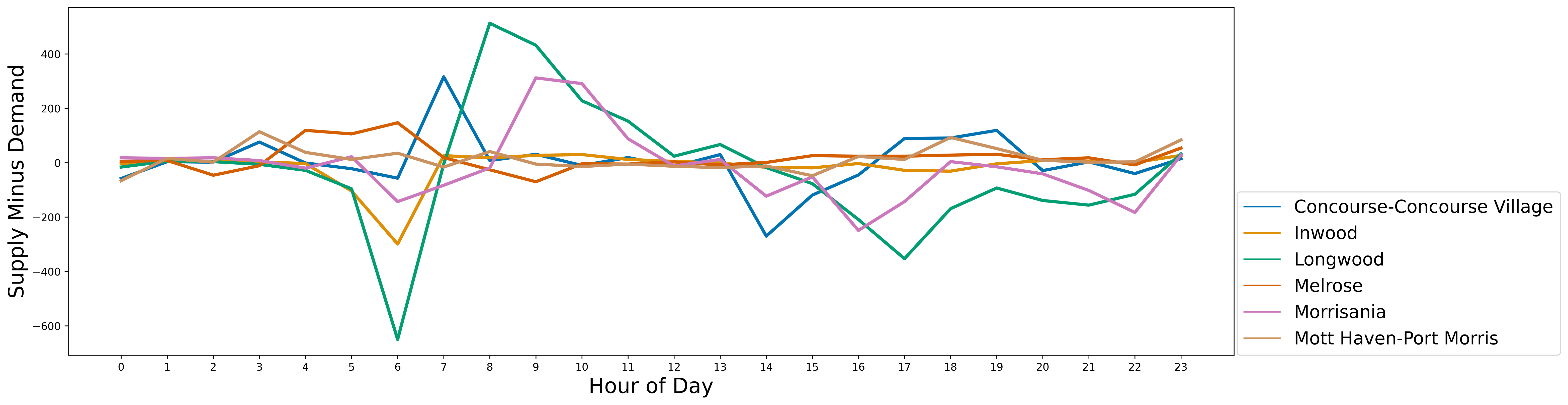}
    \caption{Bike surplus throughout the day for 6 NTAs in the expansion zone with high $\zeta$ values.}
    \label{fig:od_exp_good}
\end{figure}

\begin{figure}[H]
    \centering
    \includegraphics[width=1\linewidth]{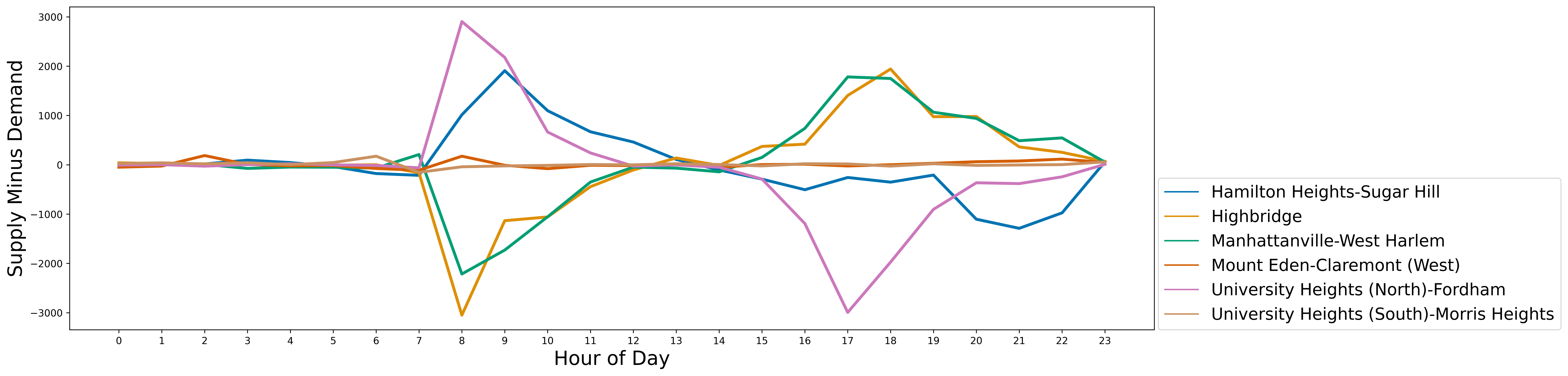}
    \caption{Bike surplus throughout the day for 6 NTAs in the expansion zone with lower $\zeta$ values.}
    \label{fig:od_exp_mid}
\end{figure}

\begin{figure}[H]
    \centering
    \includegraphics[width=1\linewidth]{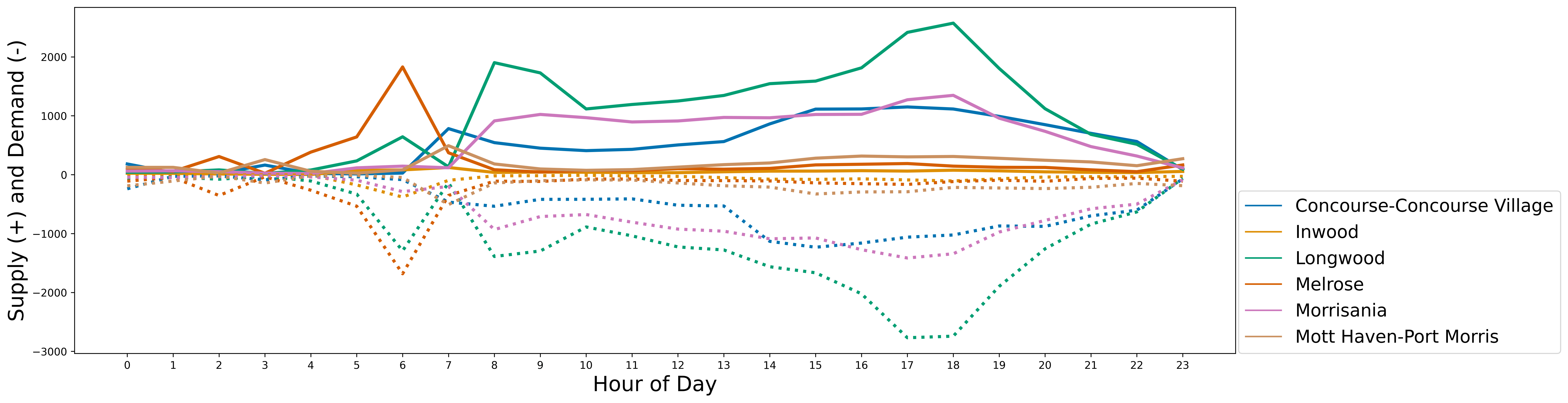}
    \caption{Bike demand and supply throughout the day for NTAs in the expansion zone with high $\zeta$ values.}
    \label{fig:od_exp_good_ds}
\end{figure}

\begin{figure}[H]
    \centering
    \includegraphics[width=1\linewidth]{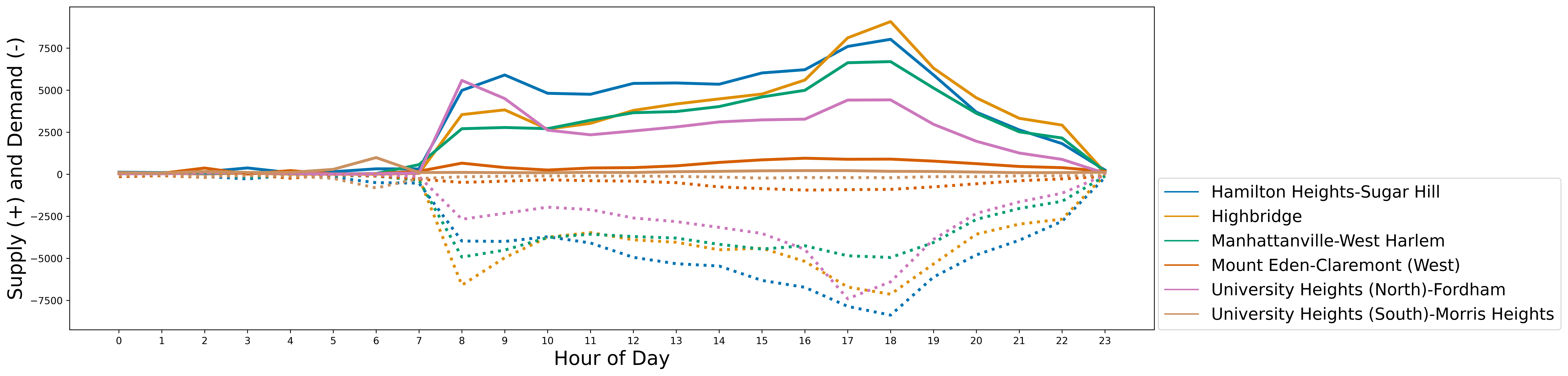}
    \caption{Bike demand and supply throughout the day for NTAs in the expansion zone with lower $\zeta$ values.}
    \label{fig:od_exp_mid_ds}
\end{figure}

We then observed the surplus in the 12 expansion zone NTAs, as shown in \Cref{fig:od_exp_good} and \Cref{fig:od_exp_mid}. The supply and demand of these 12 areas are shown in \Cref{fig:od_exp_good_ds} and \Cref{fig:od_exp_mid_ds}.



\begin{table}[H]
    \centering
    \begin{tabularx}{\textwidth}{XXXXXXXXX}
    \hline
        \textbf{NTA} & \textbf{Demand, Before, 8AM} & \textbf{Demand, After, 8AM} & \textbf{Supply, Before, 8AM} & \textbf{Supply, After, 8AM} & \textbf{Demand, Before, 6PM} & \textbf{Demand, After, 6PM} & \textbf{Supply, Before, 6PM} & \textbf{Supply, After, 6PM}\\ \hline
        Bed-Stuy (East) & 890 & 1729 & 291 & 973 & 574 & 1646 & 800 & 1933 \\ \hline
        Bed-Stuy (West) & 1723 & 1861 & 893 & 1256 & 1422 & 2415 & 1720 & 2822 \\ \hline
        Crown Heights (North) & 891 & 1978 & 338 & 1201 & 550 & 2423 & 773 & 2593 \\ \hline
        Crown Heights (South) & 112 & 602 & 22 & 446 & 94 & 872 & 129 & 916 \\ \hline
        UWS (Central) & 6082 & 3363 & 1689 & 1632 & 3594 & 2790 & 5976 & 3933 \\ \hline
        UWS-Lincoln Square & 5868 & 6054 & 2441 & 3005 & 4954 & 4789 & 6010 & 6497 \\ \hline
        UWS-Manhattan Valley & 2557 & 5769 & 533 & 3708 & 1401 & 5829 & 2375 & 7572 \\ \hline
    \end{tabularx}
    \caption{Trip and count with status, commute rushes, before and after the expansion}
    \label{tab:od_ds}
\end{table}

Focusing on the 6 NTAs with consistently high $\zeta$ values (\Cref{fig:od_exp_good}), we see the magnitude of the surplus is significantly smaller compared to the ones shown earlier in \Cref{fig:nta_od_bvsa} b) and c), indicating a well-balanced supply and demand, which may explain the high $\zeta$ values in \Cref{fig:zraw_few}. While the surplus magnitude in \Cref{fig:od_exp_mid} was comparable to those in \Cref{fig:nta_od_bvsa} b) and c), the $\zeta$ values in \Cref{fig:zraw_few} were higher overall in the expansion zones than in the UWS and the selected Brooklyn areas. Inspecting the actual usage based on \Cref{fig:od_exp_good_ds} showed that some expansion NTAs with high $\zeta$ might be explained by low usage (i.e., Inwood and Melrose), while the usage in others among these 6 NTAs had comparable usage to Bed-Stuy. \Cref{fig:od_exp_mid_ds} revealed that the 6 NTAs with relatively lower $\zeta$ values for the expansion zones had usage comparable to those in the UWS. Based on the observation of overall comparable usage and higher $\zeta$ values, the analysis above suggests that the balance between supply and demand may be slightly better in the expansion zones. 

\subsection{\textit{Service Performance Disparity, $ D_{sp}$}}
\begin{figure}[H]
    \centering
    \includegraphics[width=0.6\linewidth]{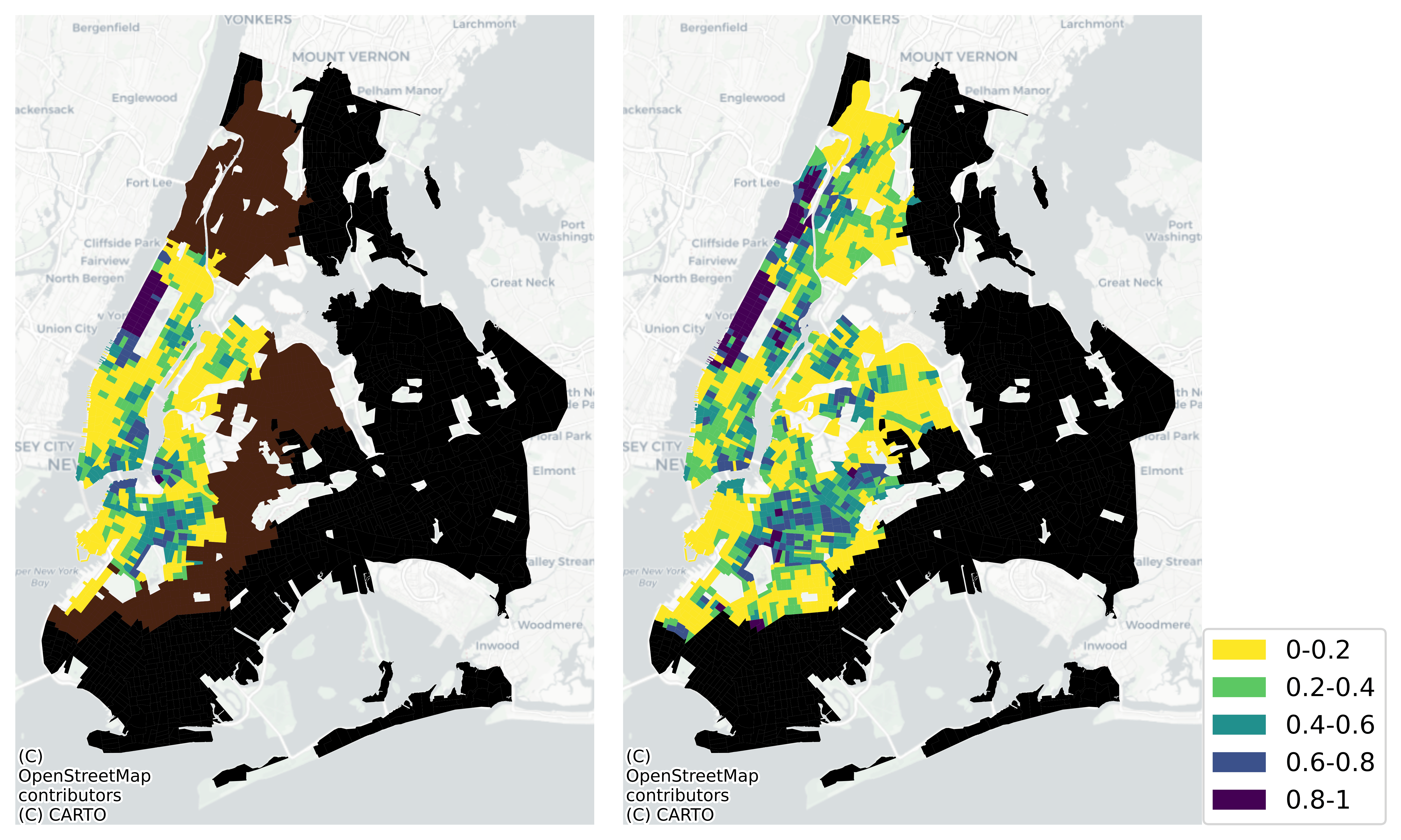}
    \caption{$D_{sp}$ values for hour 8, before (left) and after (right) the Citi Bike expansion. Brown represents tracts that do not yet have access to Citi Bike services. Black represents tracts that never received Citi Bike services.}
    \label{fig:d_sp}
\end{figure}

After examining the $\zeta$ (and consequently $z$) values, we inspected our results for $D_{sp}$. \Cref{fig:d_sp} shows an example of $D_{sp}$. This metric enhances the $z$ values (i.e., the reversed rank of $\zeta$) by involving the population count element, where a higher $D_{sp}$ value indicates more people are experiencing unsatisfactory services. As a result, we observed a higher $D_{sp}$ value than the $z$ value for some areas since we designed this metric to amplify high population counts.

\subsection{\textit{Destination Access Insufficiency, $I_{access}$}}
\begin{figure}[H]
    \centering
    \includegraphics[width=0.6\linewidth]{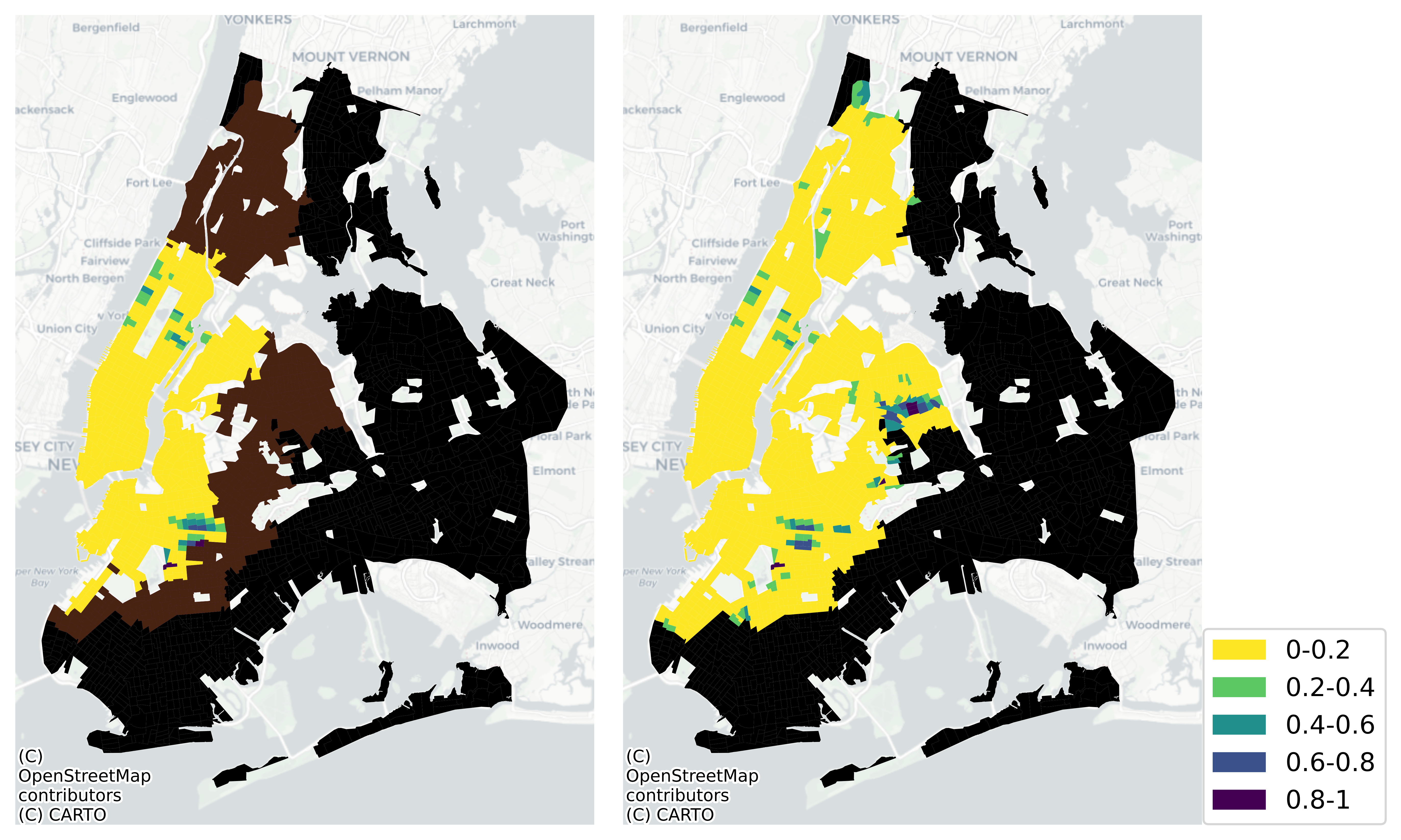}
    \caption{$I_{access}$ for commute zones before (left) and after the Citi Bike expansion (right). Brown represents tracts that do not yet have access to Citi Bike services. Black represents tracts that never received Citi Bike services.}
    \label{fig:iacc_com}
\end{figure}

\begin{figure}[H]
    \centering
    \includegraphics[width=0.6\linewidth]{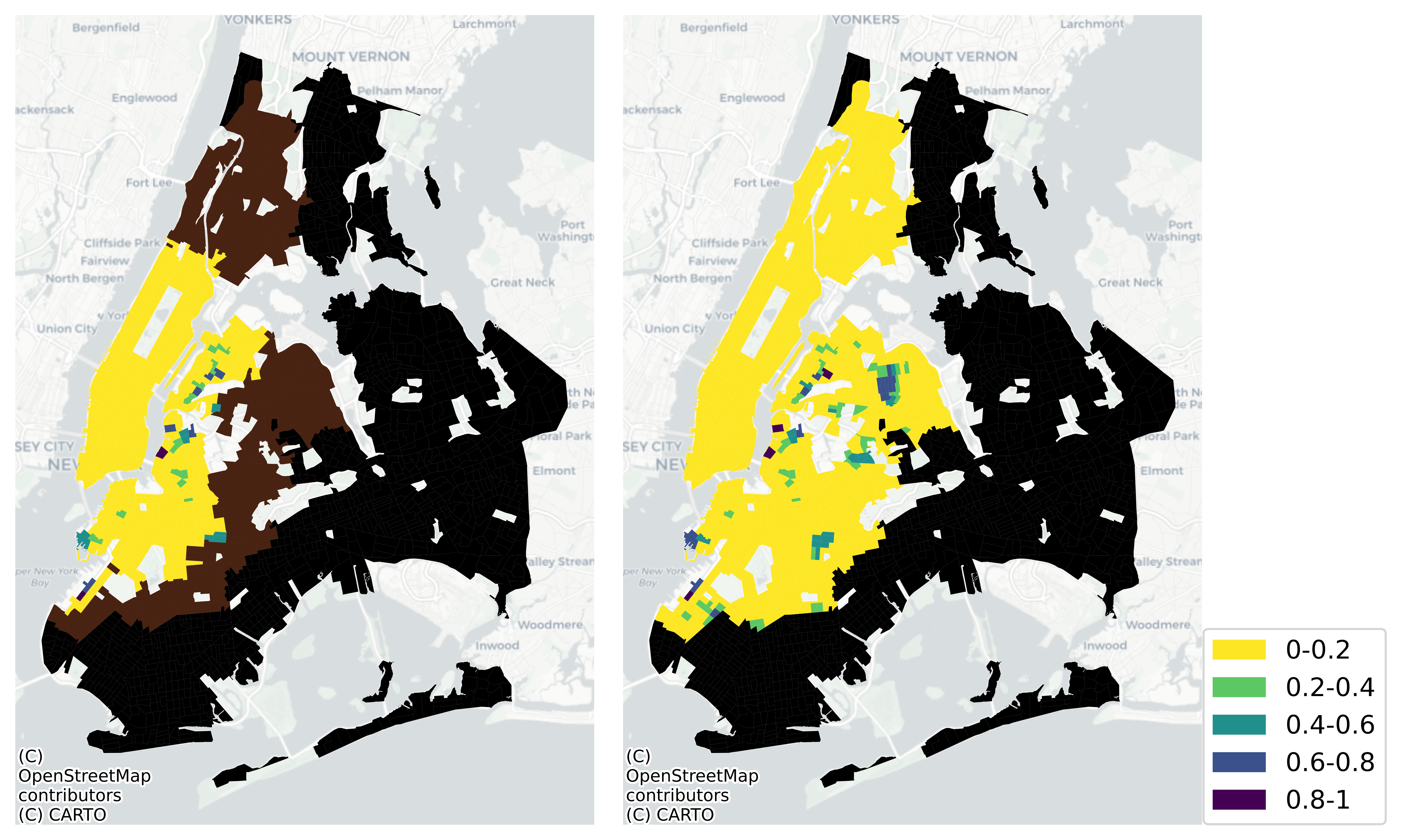}
    \caption{$I_{access}$ for recreational zones before (left) and after the Citi Bike expansion (right). Brown represents tracts that do not yet have access to Citi Bike services. Black represents tracts that never received Citi Bike services.}
    \label{fig:iacc_rec}
\end{figure}

\Cref{fig:iacc_com} and \Cref{fig:iacc_rec} show the $I_{access}$ results for commute and recreational zones. A lower value indicates better accessibility and vice versa. Generally, the majority of the study area has a good $I_{access}$ score, indicating that the density and distribution of the Citi Bike stations are well-designed.

The areas with high $I_{access}$ commute zone scores are primarily residential-heavy, while those with high recreational zone scores are tracts relatively far away from recreational activities such as parks and shopping centers. While a higher $I_{access}$ may indicate more people are experiencing relatively worse destination reachability, it is worth noting that we used the city’s convoluted zoning data, which primarily consisted of mixed zones, instead of identifying job centers using job-specific data such as the Longitudinal Household-Employer Data.

\subsection{\textit{Performance-Integrated Destination Access Insufficiency, $ZI_{access}$}}
\begin{figure}[H]
    \centering
    \includegraphics[width=0.6\linewidth]{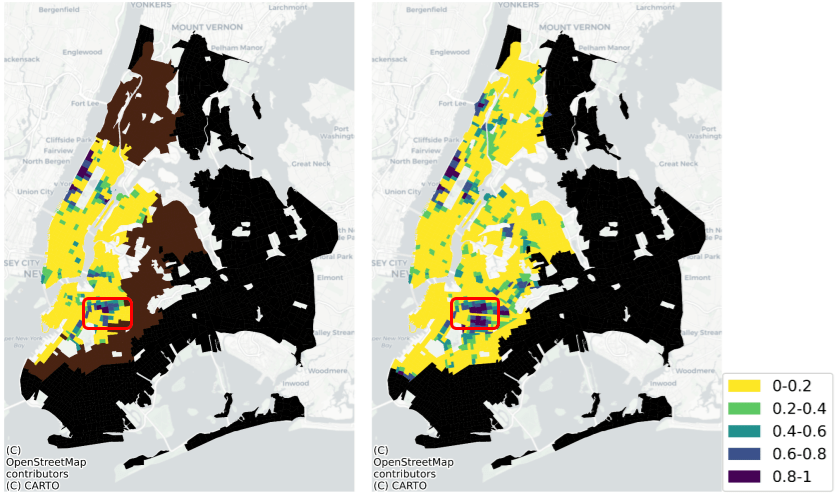}
    \caption{$ZI_{access}$ for commute zones, hour 8, before (left) and after the Citi Bike expansion. Brown represents tracts that do not yet have access to Citi Bike services. Black represents tracts that never received Citi Bike services. Red boxes highlight the Bed-Stuy area.}
    \label{fig:ziacc_com}
\end{figure}

\Cref{fig:ziacc_com} shows an example of $ZI_{access}$, the proposed multi-dimensional, high-granularity equity index before and after the Citi Bike expansion, where a higher score indicates more people are experiencing insufficient destination access or unsatisfactory service performance. For instance, focusing on the Bed-Stuy area (red box in \Cref{fig:ziacc_com}), we observed worsened $ZI_{access}$ scores after the expansion, potentially due to decreased service availability, which was also reflected in the observations from \Cref{fig:nta_od_bvsa} c) and \Cref{tab:od_ds}.

\begin{figure}[H]
    \centering
    \includegraphics[width=0.6\linewidth]{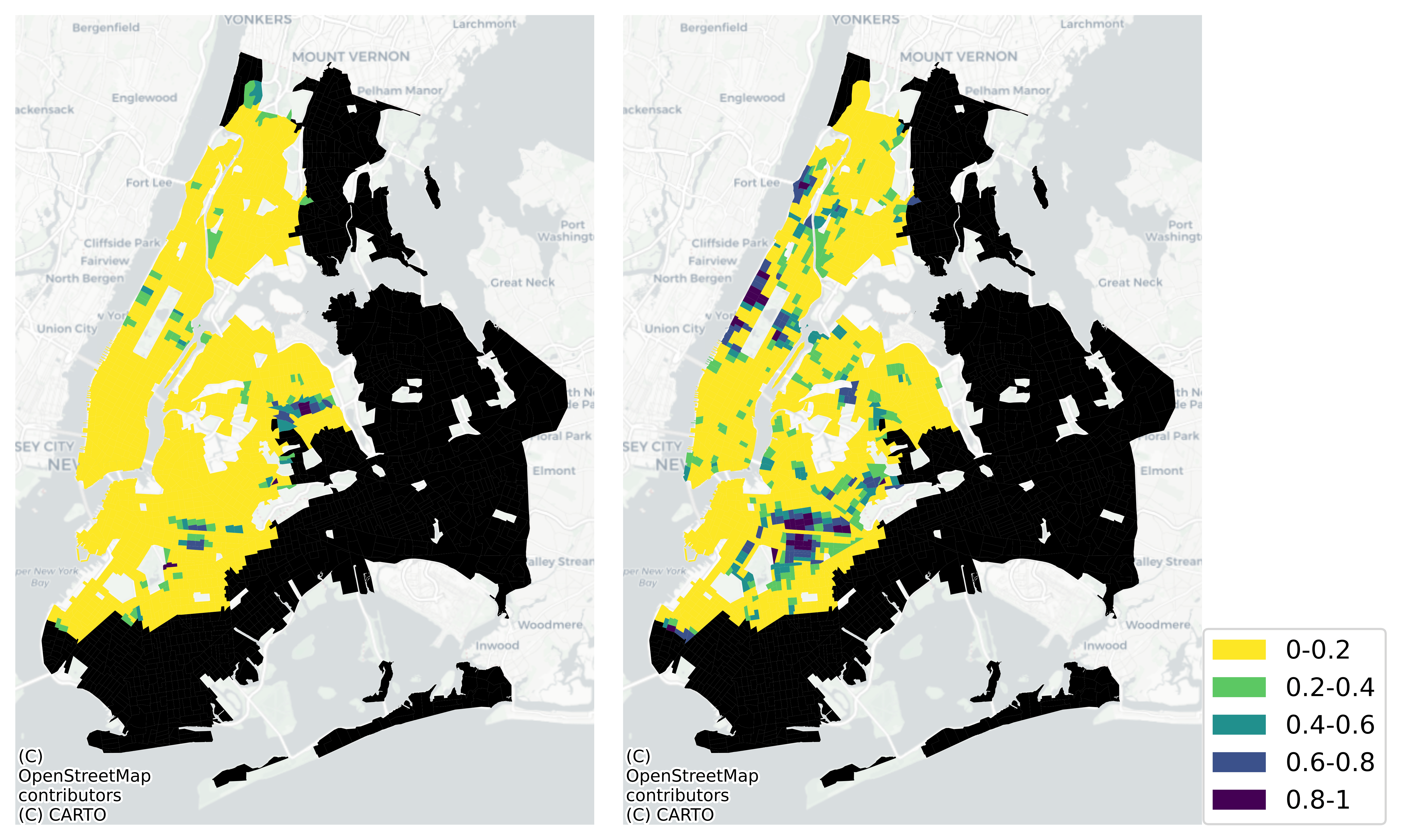}
    \caption{$I_{access}$ (left) and $ZI_{access}$ at hour 8 (right) for commute zones, after the Citi Bike expansion. Brown represents tracts that do not yet have access to Citi Bike services. Black represents tracts that never received Citi Bike services.}
    \label{fig:zivsi_com}
\end{figure}

\begin{figure}[H]
    \centering
    \includegraphics[width=0.6\linewidth]{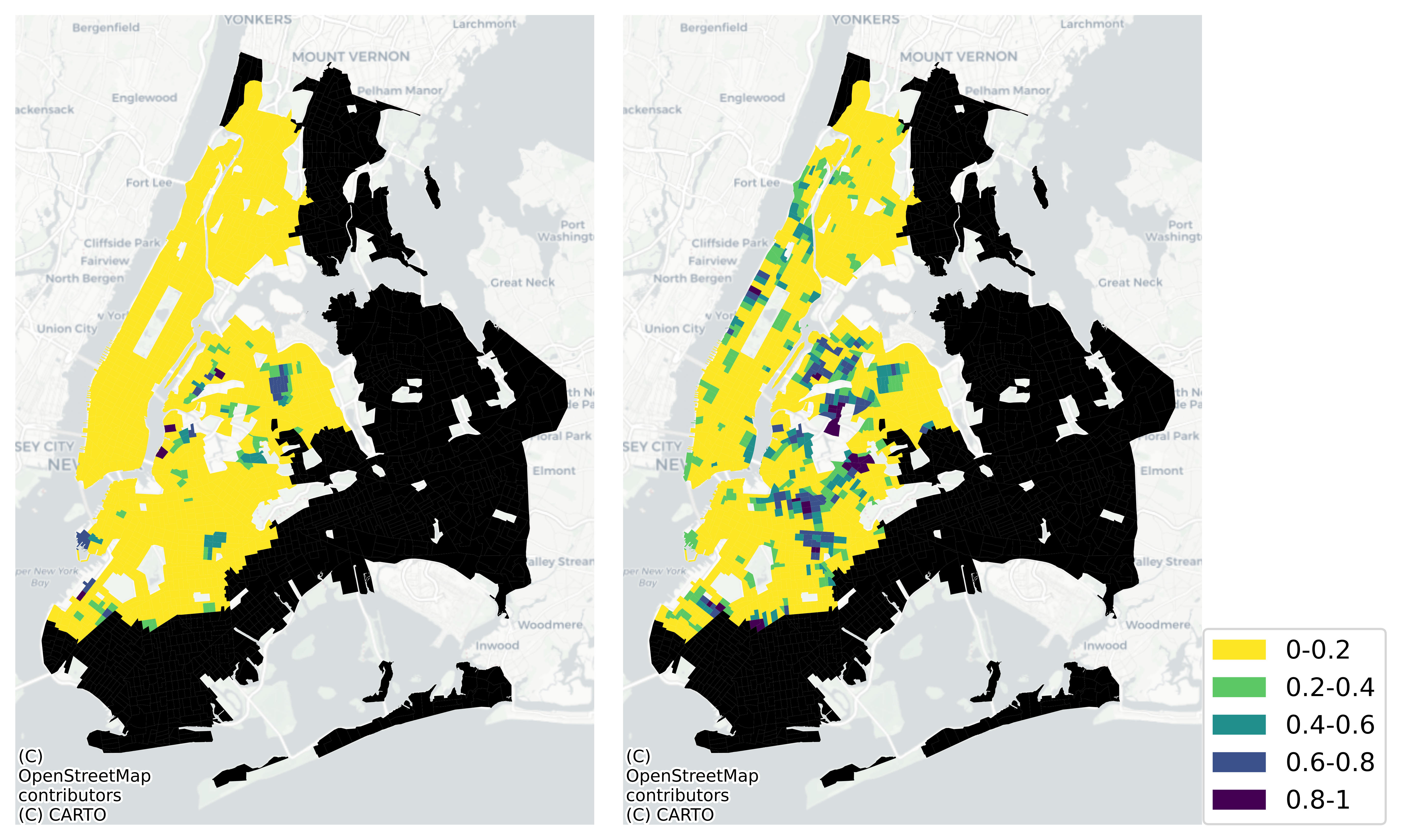}
    \caption{$I_{access}$ (left) and $ZI_{access}$ at hour 8 (right) for recreational zones, after the Citi Bike expansion. Brown represents tracts that do not yet have access to Citi Bike services. Black represents tracts that never received Citi Bike services.}
    \label{fig:zivsi_rec}
\end{figure}

\Cref{fig:zivsi_com} and \Cref{fig:zivsi_rec} are examples of comparing $ZI_{access}$ and $I_{access}$ after the Citi Bike expansion. As expected, we observed higher $ZI_{access}$ compared to $I_{access}$ for areas with low service quality, highlighting the effects of integrating service performance quality into destination accessibility analysis. The difference between the $I_{access}$ and $ZI_{access}$ results emphasizes that sufficient accessibility is not guaranteed when service availability is unsatisfactory, further affirming the importance of considering quality when assessing accessibility by different services. 

While the calculated equity index pinpoints locations that may need service or accessibility improvement, practitioners often consult predefined geographical boundaries such as the Neighborhood Tabulation Areas in New York City. However, while traditional geographical boundaries may sometimes be enough to represent the needs of the population within each area, often, the variations within these predefined boundaries warrant a more nuanced approach to identifying boundaries to better connect locations with similar needs. 

Given the nature of our study, we would use the calculated equity index and the demographic information for the analyzed census tracts as a proxy to identify similar needs from the populations in each tract. With the aforementioned motivation in mind, we sought methods to create boundaries based on various attributes attached to each census tract. The results of this analysis should distinguish differences between a set of predefined geographical boundaries and the new set of boundaries created based on the calculated equity index and the demographic attributes from the synthetic population. Specifically, we aimed to arrive at a method where the input is a point layer that includes the calculated equity index and available demographic information in the synthetic population data, such as income, race, and English proficiency, and the output is the structural representation of all the points in the input layer that identifies points with similar attributes. In addition to better incorporating the demographic data from the synthetic population dataset, the results from the structural representations of the input points enhance our proposed framework from an evaluation tool to an evaluation-screening tool by allowing users to further identify populations and areas with specific demographic characteristics and require additional improvement in transportation equity. 

To achieve the objectives mentioned above, we utilized the GraphSAGE model available through the StellarGraph python package \cite{stellargraph}, which can learn graph embeddings from every node in a certain graph and is especially useful for graphs with rich node attribute information. After learning the relationship between all the points in the input layer based on all the relevant attributes including the calculated equity index and demographic information), we perform a clustering analysis on the embedded input layer to compare with NTA, a traditionally used geographic in NYC. By pinpointing the differences between the clustered embeddings from the GraphSAGE output and the geographical boundaries that each input point traditionally falls into, our methodology showcases how decision-makers may achieve targeted improvements for transportation equity. 

To demonstrate how we can use GraphSAGE with our proposed framework, we selected the centroid of each of the 1117 census tracts in our analysis as the location for our input nodes, and each node carries the following attributes: the calculated equity score, total population count, total household, non-white population, limited-English population, low-income household, and households with no vehicle ownership. For this discussion, we focus on the embedding and clustering for the stationary $I_{access}$ layer and 5 $ZI_{access}$ layers with 5-hour increments starting from 3 AM. To best accompany the clustering objectives of this analysis, we utilized an unsupervised GraphSAGE model as described by  \cite{stellargraph}. We trained the embedding model for 100 epochs for each of the six input layers and achieved accuracy scores of approximately 0.85 across all layers. The accuracy score was calculated as the ratio of the total number of correct predictions (true positives + true negatives) to the total number of predictions (true positives + true negatives + false positives + false negatives).


Once we obtained the embeddings of these centroid nodes from the trained GraphSAGE models, we utilized k-means clusterings to arrive at 128 clusterings based on the embeddings. By choosing the same number of clusters for the embeddings and the unique number of NTAs in our study area, we were able to compare the differences between the learned boundaries based on the list of attributes and the predefined boundaries based on geographic proximity. We compared the clusters from the embeddings to the NTA designation to showcase the potential of using different attributes to better target population segments with different characteristics. In addition to comparing overall differences between node embedding clustering and NTA, we also compare the clustered differences between the layer stationary $I_{access}$ and the 5 temporally dynamic MDHG equity index ($ZI_{access}$) layers. As such, we further demonstrate the dynamic component of the proposed MDHG framework and can potentially draw further policy implications according to the temporal variations and needs of various population segments. Since the input to the graph representation learning model is flexible, practitioners may opt to change the detail level for the node (e.g., use person-level nodes instead of tract-level nodes) and the attributes to consider in the model. 

\begin{figure}[H]
    \centering
    \includegraphics[width=0.8\linewidth]{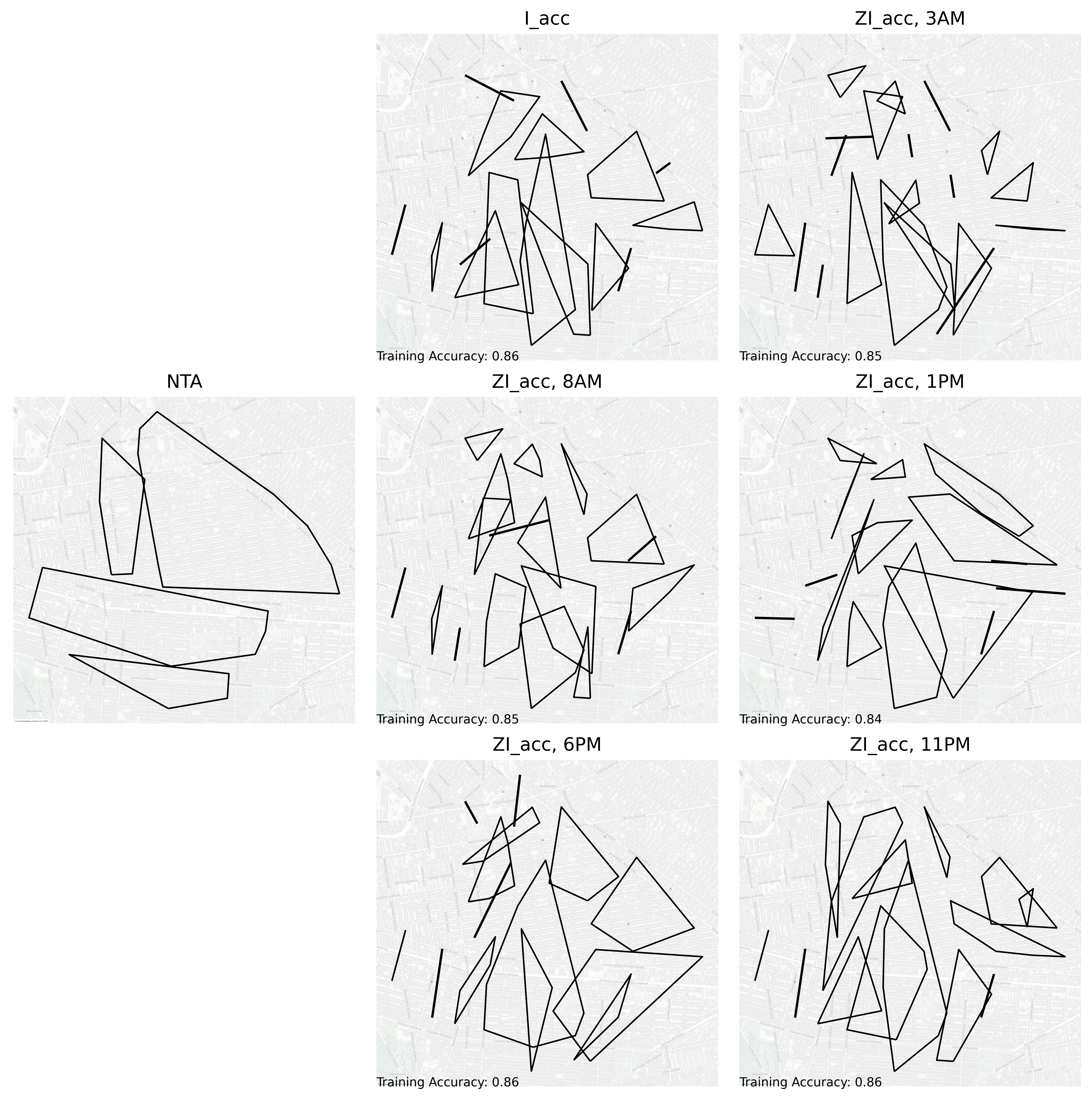}
    \caption{NTA compared with clustering results for census tracts in Bed-Stuy and Crown Heights.}
    \label{fig:gsage_nta_vs_clus}
\end{figure}

We use the shape of the clusters or NTA, represented using polygons, to compare the result of the clustering label from the embeddings and the NTA label for each of the 1117 tracts in our study area. Using Bed-Stuy and Crown Heights as an example, \Cref{fig:gsage_nta_vs_clus} shows that the clusters from the node embeddings using $I_{access}$ or $ZI_{access}$ and demographic attributes are different than the NTA labels, proving that clusters based on various attributes show a more nuanced picture of the different experiences in different communities. In addition, the differences between the $ZI_{access}$ layers and the $I_{access}$, plus the variations among the $ZI_{access}$ layers, further demonstrate the dynamic component of the proposed MDHG equity index. The results from this analysis suggest that practitioners should consider using learned boundaries based on their desired attributes to address issues in different communities.


\subsection{\textit{Service Recovery Rate, $SR$}}
\begin{figure}[H]
    \centering
    \includegraphics[width=1\linewidth]{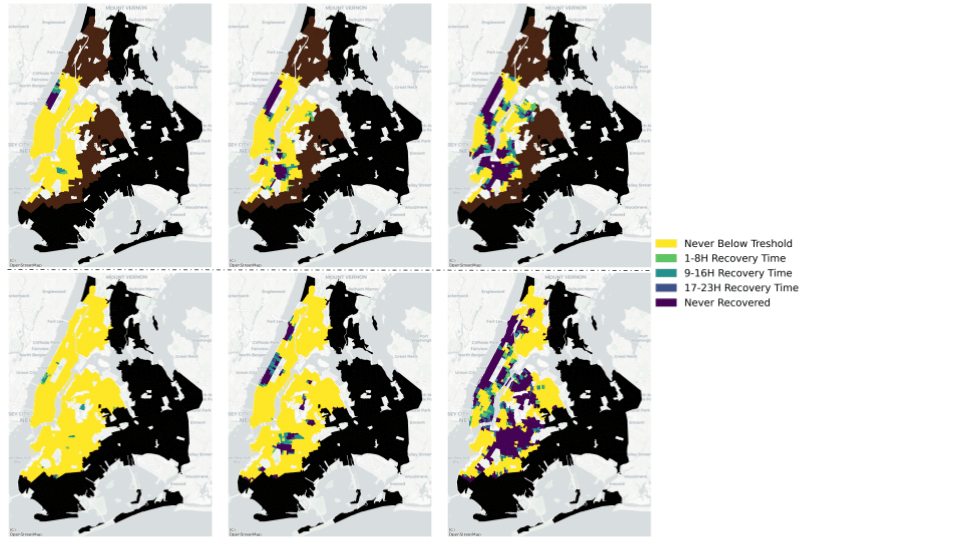}
    \caption{$R_l$ for l=[0.4, 0.6, 0.8] (left to right), before (above dashed line) and after the Citi Bike expansion (below dashed line). Brown represents tracts that do not yet have access to Citi Bike services. Black represents tracts that never received Citi Bike services.}
    \label{fig:recov_time}
\end{figure}

\Cref{fig:recov_time} shows the recovery time ($R$) for each tract responding to the six $\zeta$ threshold levels. As the threshold level increases, more tracts take longer to recover to the specified level. By comparing the results before and after the expansion, we observed fewer tracts falling under the lower thresholds while more tracts went below the higher thresholds after the expansion, thus echoing the previous observations of more balanced bike availability. While the $\zeta$ values from earlier provide the hourly change in service availability, the $R$ values provide a comprehensive representation of the service reliability across all time periods inspected (e.g., the 24-hour period in each day for this case study). The calculation's configuration also allows practitioners to customize the target threshold for service availability based on their definition of ideal service reliability to pinpoint communities that may need improvement to reach a desired service availability. 

\begin{figure}[H]
    \centering
    \includegraphics[width=1\linewidth]{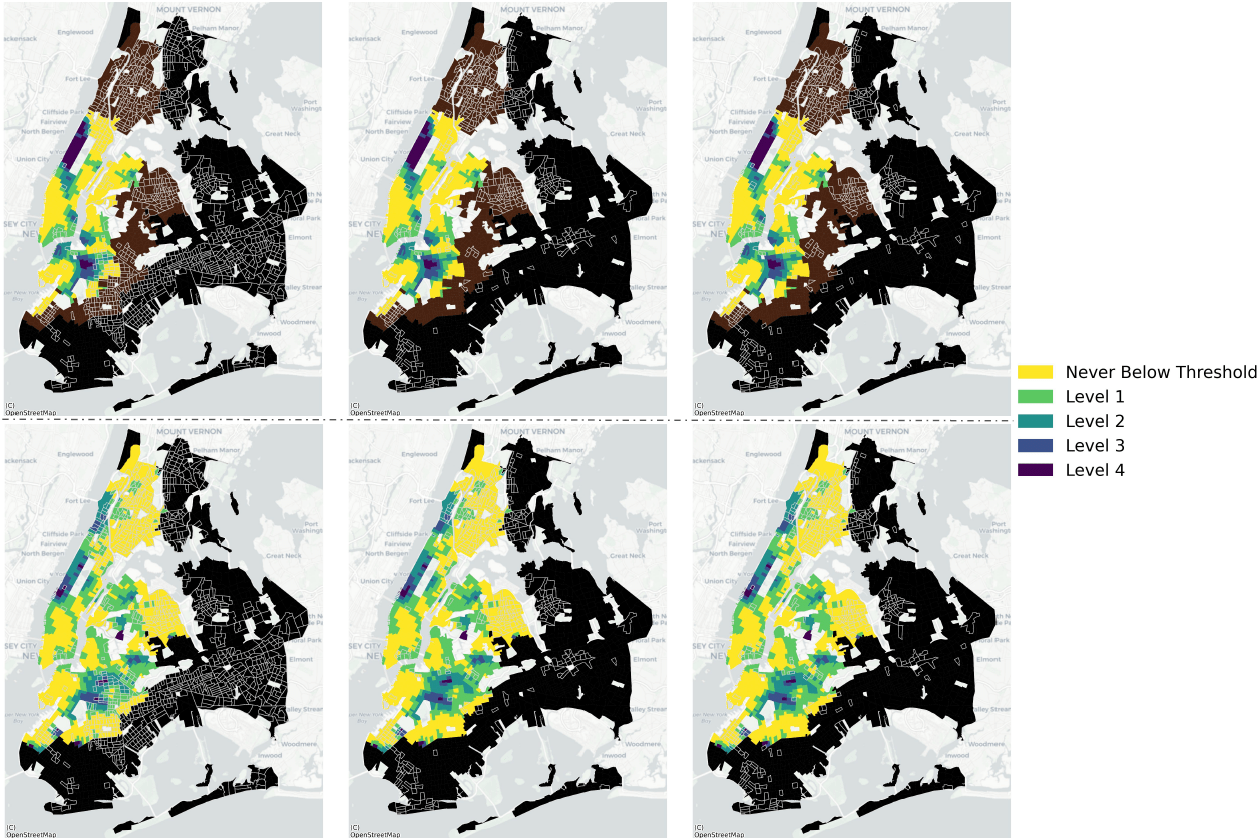}
    \caption{$SR$ values with three demographic overlays, before (above dashed line) and after the Citi Bike expansion (below dashed line). Light gray lines represent minority neighborhoods. Brown represents tracts that do not yet have access to Citi Bike services. Black represents tracts that never received Citi Bike services.}
    \label{fig:sr_integral}
\end{figure}

After calculating $SR$ using \Cref{eq:SR}, we created five categories from this integrated value, where tracts with a $SR$ of 0 were considered to never reach below the threshold, while the other $SR$ values, which ranged from slightly above 0 to 12, were categorized based on increments of 3. In other words, the higher the $SR$ level, the longer it takes a tract to recover above the $\zeta$ threshold. \Cref{fig:sr_integral} shows the $SR$ values with three population overlays: 1) minority neighborhoods (percentage of non-white residents was above the 75 percentile), 2) areas with a relatively high percentage of the population who are of minority races, below the poverty line, and speak limited English, and 3) adding two additional criteria, do not own vehicle and unemployment, to the three conditions in the previous layer. As shown in \Cref{fig:sr_integral}, the Bronx (i.e., northmost of the expansion zone) had the highest number of combination demographic tracts, indicating the overall area as disadvantaged based on these population segments. However, the $SR$ values after the expansion show that the majority of the tracts in the Bronx were always above all of the tested thresholds, thus indicating consistently high service availability. This is mostly not due to under-utilization, as \Cref{fig:od_exp_good_ds} and \Cref{fig:od_exp_mid_ds} from an earlier section showed that many areas in the expansion zones have comparable usage compared to areas with Citi Bike service before the expansion. Table 5.2 shows a few statistics of the SR values comparing the disadvantaged populations with the rest after the expansion. We included the mean value ($\mu$) for each population group and performed t- and ks-tests to compare the mean and distribution of the disadvantaged and not disadvantaged populations.

\Cref{tab:sr_stats} shows the mean $SR$ values after the expansion are significantly lower for all inspected disadvantaged population groups than the rest, and have a different distribution. By comparing the $SR$ values before and after the expansion, we witnessed an overall improvement in recovery time to the selected thresholds, suggesting a generally improved Citi Bike system rebalance. Nevertheless, this observation needs to be considered concurrently with a previous one, in which we saw worsened bike availability in certain neighborhoods while others saw improvement.

\begin{table}[H]
    \centering
    \begin{tabularx}{\textwidth}{XXXXX}
    \hline
        \textbf{Combination} & \textbf{$\mu$, Disadvantaged} & \textbf{$\mu$, Not Disadvantaged} & \textbf{$p$, t-test} & \textbf{$p$, ks-test} \\ \hline
        Race & 1.3106 & 1.8362 & 0.0 & 0.0 \\ \hline
        Race, Poverty, Limited English & 0.7889 & 1.5346 & 0.0 & 0.0 \\ \hline
        Race, Poverty, Limited English, No Vehicle, Unemployed & 0.8460 & 1.5197 & 0.0 & 0.0 \\ \hline
    \end{tabularx}
    \caption{Trip and count with status, commute rushes, before and after the expansion}
    \label{tab:sr_stats}
\end{table}

\subsection{\textit{Summary of Results}}
We applied the proposed equity index framework to the Citi Bike expansion case study. We first used the $\zeta$ values in the service performance metric to explore temporal variations in bike availability across the study area. While the service availability generally improved after the expansion, we identified certain neighborhoods that consistently had low service availability before and after the expansion. We applied the $D_{sp}$ methodology to consider population count, where a higher $D_{sp}$ value indicates more people are experiencing unsatisfactory service. The $I_{access}$ score highlighted areas with insufficient access to commute or recreational activity zones. After combining the service performance metric with the destination accessibility insufficiency, we obtained the people-oriented MDHG equity index, $ZI_{access}$, where a higher metric value indicates more people are experiencing low service availability or insufficient access. We observed worsened $ZI_{access}$ scores for areas with lower service availability after the expansion, thus highlighting the effect of this metric in involving service availability. Furthermore, the service recovery rate metric $SR$ showed overall improvement in recovering to different threshold levels after the expansion. We first noticed that many disadvantaged areas were in the expansion zone by looking at the disadvantaged neighborhoods based on various demographic combinations. The $SR$ values for these neighborhoods were primarily 0, indicating that the service availability in these areas was consistently high. 

\section{Implications}
The findings from the case study reveal several important policy implications and recommendations. It is essential to consider the varying levels of service quality when evaluating transportation services. While accessibility to service facilities is crucial, it does not necessarily guarantee excellent service quality. Transportation projects, especially service-based ones, often exhibit fluctuations in service levels over time. For instance, the performance metric $z$ in our case study highlighted the spatial-temporal variations in bike availability. Therefore, policymakers should evaluate transportation services by considering not only service and destination accessibility but also the fluctuating levels of service quality. This comprehensive approach will provide a more accurate assessment of the effectiveness of transportation services.

Furthermore, the analysis demonstrated that areas with similar equity metrics might have different demographic characteristics, reflecting distinct needs among residents. For example, two neighborhoods might experience inadequate service availability and insufficient destination accessibility in a bike-share program. However, one neighborhood may have the financial resources to opt for alternative transportation, while the other may not, resulting in a higher burden on the latter. To address these disparities, it is recommended that policymakers use high-granularity demographic data, such as synthetic population datasets, to identify specific population segments by combining various demographic attributes. This detailed demographic analysis will enable a more targeted and effective allocation of resources and interventions, ensuring that the needs of different communities are met more accurately.

Moreover, incorporating spatial-temporal data, including detailed, person-level trip data, can significantly enhance the understanding of stakeholders' unique needs. Future studies should include such data to extend the analysis by capturing activity patterns and providing deeper insights into individuals' daily movements and transportation needs. By doing so, policymakers can design more tailored and responsive transportation services that better meet the diverse requirements of the population.

Based on these findings, policymakers can leverage the performance-incorporated accessibility index as a tool to assess service quality over time, ensuring that service levels are consistently maintained. The index can also help target specific population segments more accurately by utilizing high-granularity demographic data, allowing for more equitable and effective resource allocation.

\section{Limitation and Future Work of the Study}
It is worth mentioning a few limitations. For service availability, we assigned rows with bike availability below 0 or over the tract’s capacity to 0 since we encountered these cases after filtering the trip data. This step might have affected the accuracy of the $\zeta$ values. In addition, we selected data for one month each to represent before and after the CB expansion, which might limit our observations. Future works should consider using trip data covering a more extended period for more prolonged analysis. Furthermore, the 2600m buffer distance from each home tract might not have been enough since it was Euclidean. Instead of using zoning data, higher-resolution points of interest data with specified origins would improve the $I_{access}$ analysis.

\section{Conclusions}
This study identified the need for equity-focused project evaluation tools and addressed the gap in combining destination accessibility with service quality by proposing a multi-dimensional, high-granularity (MDHG) equity index. Applying this framework to a Citi Bike expansion case study provided key insights into transportation equity. The results showed improved service quality in some areas (e.g., Upper West Side) but declines in others (e.g., Bedford-Stuyvesant). The MDHG equity score, $ZI_{access}$, highlighted areas with poor service levels and limited access, while the Service Recovery metric, $SR$, indicated overall system reliability improvements. Our analysis demonstrated that different population groups experience varying levels of service availability and accessibility. Policy implications suggest that accessibility alone does not guarantee excellent service quality, as shown by performance metric $z$. Thus, evaluating both service and destination accessibility alongside service quality is crucial. Additionally, areas with similar equity metrics may have different demographic characteristics, requiring targeted resource allocation based on high-granularity demographic data. Despite limitations, such as the lack of high-resolution points of interest data, our framework emphasizes the importance of including service quality and detailed demographics in transportation planning. The MDHG equity index provides a comprehensive approach to assessing transportation equity, improving planning, and ensuring more equitable service delivery for diverse populations. Future work should integrate person-level synthetic activity data to enhance analysis accuracy.

\section{Acknowledgements}
The contents of this paper reflect the views of the authors, who are responsible for the facts and the accuracy of the information presented herein. The work is funded, partially or entirely, by the SEMPACT center under Grant Number 69A3552348302 and the C2SMARTER Center under Grant Number 69A3552348326 from the U.S. Department of Transportation's University Transportation Centers Program. However, the U.S. Government assumes no liability for the contents or use thereof.

\section{Author Contributions}
The authors confirm their contribution to the paper as follows: study conception and design: Zilin Bian, Mengke (Enola) Ma; data curation: Mengke (Enola) Ma, Hai Yang; formal analysis: Mengke (Enola) Ma, Zilin Bian; methodology: Mengke (Enola) Ma, Zilin Bian; software: Mengke (Enola) Ma; visualization: Mengke (Enola) Ma; writing - original draft: Mengke (Enola) Ma, Zilin Bian, Hai Yang, Jingqin  Gao, Joseph Chow, Kaan Ozbay; writing - review \& editing: Mengke (Enola) Ma, Zilin Bian, Jingqin  Gao, Joseph Chow, Kaan Ozbay. All authors reviewed the results and approved the final version of the manuscript.

\bibliography{refs} 
\bibliographystyle{unsrt}
\end{document}